\documentclass[modern]{aastex63}

\usepackage{graphics,graphicx,mhchem, courier,lineno} 

\usepackage{natbib}
\bibpunct{(}{)}{;}{a}{,}{,}

\newcommand{\GG}[1]{}
\usepackage[symbol]{footmisc}

\defcitealias{lellouch_19}{L19}

\begin{document}

\title{Variability in Titan's Mesospheric HCN and Temperature Structure as Observed
  by ALMA}

\correspondingauthor{Alexander E. Thelen}
\email{alexander.e.thelen@nasa.gov}

\author{Alexander E. Thelen}
\affiliation{Solar System Exploration Division, NASA Goddard Space
  Flight Center, Greenbelt, MD 20771, USA}

\author{Conor A. Nixon}
\affiliation{Solar System Exploration Division, NASA Goddard Space Flight Center, Greenbelt, MD 20771, USA}

\author{Richard Cosentino}
\affiliation{Space Telescope Science Institute, Baltimore, MD 21218, USA}

\author{Martin A. Cordiner}
\affiliation{Solar System Exploration Division, NASA Goddard Space
 Flight Center, Greenbelt, MD 20771, USA}
\affiliation{Department of Physics, Catholic
  University of America, Washington, DC 20064, USA}

 \author{Nicholas A. Teanby}
\affiliation{School of Earth Sciences, University of Bristol, Bristol BS8 1RJ, UK}

\author{Claire E. Newman}
\affiliation{Aeolis Research, Chandler, AZ 85224, USA}

 \author{Patrick G. J. Irwin}
\affiliation{Atmospheric, Oceanic and Planetary Physics, Clarendon Laboratory, University of Oxford, Oxford OX1 3PU, UK}

\author{Steven B. Charnley}
\affiliation{Solar System Exploration Division, NASA Goddard Space Flight Center, Greenbelt, MD 20771, USA}

\renewcommand{\thefootnote}{\fnsymbol{footnote}}
\setcounter{footnote}{1}

\footnotetext{NASA Astrobiology Postdoctoral Program Fellow under the
  management of the Universities Space Research Association and Oak
  Ridge Associated Universities.}

\begin{abstract}
The temperature structure of Titan's upper atmosphere exhibits large
variability resulting from numerous spatially and temporally irregular
external energy sources, seasonal changes, and the influence of
molecular species produced via photochemistry. In particular, Titan's
relatively abundant HCN is thought to provide substantial cooling to
the upper atmosphere through rotational emission, balancing
UV/EUV heating and thermal conduction. Here, we present the analysis
of ALMA observations of Titan from 2012, 2014, 2015, and 2017,
corresponding to planetocentric solar longitudes of
$\sim$34--89$^{\circ}$, including vertical HCN and temperature
profiles retrieved from the lower mesosphere through the thermosphere
($\sim350$--1200 km; $3\times10^{-2}$--$2\times10^{-8}$
mbar). Throughout the atmosphere, temperature profiles differ by 10 to
30 K between observations approximately one Earth year apart,
particularly from 600--900 km. We find evidence for a large imbalance
in Titan's upper atmospheric energy budget between 2014 and 2015,
where the mesospheric thermal structure changes significantly and
marks the transition between a mesopause located at $\sim600$ km
($2\times10^{-4}$ mbar) and at $\sim$800 km ($3\times10^{-6}$ mbar). The retrieved HCN abundances vary dramatically during the 2012 to 2017 time period as well, showing close to 2 orders of magnitude difference in abundance at 1000 km. However, the change in HCN abundance does not appear to fully account for the variation in mesospheric temperatures over the $L_S\sim$34--89$^{\circ}$ period. These measurements provide additional insight into the variability of Titan's mesospheric composition and thermal structure following its 2009 vernal equinox, and motivate continued investigation of the origins of such rapid changes in Titan's atmosphere throughout its seasonal cycle. 

\end{abstract}

\section{Introduction} \label{sec:intro}
\renewcommand{\thefootnote}{\arabic{footnote}}
\setcounter{footnote}{0}
Titan, Saturn's largest moon, possesses a substantial atmosphere that
is characterized by a thermal structure similar to that of the Earth,
though colder and more vertically extended (up to $\sim1500$ km)
resulting from its
comparatively reduced insolation and low gravity
(see \citealp{flasar_14}, \citealp{yelle_14}, \citealp{horst_17}, and references therein). Through observations by the
Voyager 1, 2 and Cassini
spacecraft, and direct measurements by the Huygens probe during its
descent in 2005, Titan's atmospheric behavior and dynamics have been well
studied through roughly half of its $\sim$29.5 yr seasonal
cycle (see, for example: \citealp{smith_82, lindal_83, vervack_04, flasar_05, fulchignoni_05,
  shemansky_05, yelle_06, teanby_08b, teanby_12, snowden_13,
  vinatier_15, teanby_17, teanby_19, sylvestre_20, creecy_21,
  seignovert_21, sharkey_21}). Additionally,
measurements from ground-based observatories have helped to bridge
gaps between missions to the Saturnian system (e.g. \citealp{kim_00, geballe_03,
  gurwell_04, moreno_05}). However, the nature of short-term
variability in the dynamical state of Titan's upper
atmosphere, the seasonal evolution of its stratosphere ($\sim80$--300 km), and the
contribution of many different chemical and dynamical components to
Titan's atmospheric radiative budget still remains enigmatic after the end of the Cassini/Huygens mission. 

Models of Titan's upper atmosphere --- consisting of the mesosphere (typically defined as $\sim300$--600 km), thermosphere
($\sim600$--1000 km) and ionosphere (\textgreater1000 km) --- attempt to
simulate Titan's atmospheric thermal structure through the 
  balance of heating by solar EUV/UV photons, electron and ion 
  precipitation, interactions with Saturn's magnetosphere and plasma sheet, thermal
  conduction, and cooling by rotational and vibrational emission from
  hydrogen cyanide (HCN) and various hydrocarbons produced via photochemistry \citep{yelle_91,
    muller_00, muller_08, yelle_08, 
    bell_10, bell_11, westlake_11, snowden_14, bell_14, 
    snowden_21}. Additionally, atmospheric wave activity induced by
  gravitational tides may produce significant temperature
  perturbations in the upper atmosphere \citep{strobel_06}. These predictions, along with inferences of
temperature from \textit{in situ} density profiles obtained with the
Cassini Ion and Neutral Mass Spectrometer (INMS),
Huygens Atmospheric Structure Instrument (HASI), and measurements
derived through stellar occultations from the ground and the Cassini Ultraviolet
Imaging Spectrograph (UVIS), revealed large variability in Titan's middle and upper
atmospheric thermal structure on both diurnal and seasonal
timescales. Further, obsevations from the Cassini
  Composite Infrared Spectrometer (CIRS), Imaging Science Subsystem (ISS), Visible and Infrared Mapping Spectrometer (VIMS) instruments covering
  roughly half a Titan year indicated that Titan's radiative energy budget may
  not be fully balanced on both temporal (seasonal) and spatial
  (hemispheric) scales 
  \citep{li_15, creecy_19, creecy_21}, though further observations are
  required to determine the nature and effects of this imbalance on
  timescales of a Titan year or more.

Vertical perturbations of $\sim10$--30 K on $\sim50$--100 km
scales (on order 1--2 scale heights) were present in the
Huygens/HASI thermospheric measurements, indicating strong wave activity
\citep{fulchignoni_05, aboudan_08}. Ionospheric measurements with the INMS
revealed 30--60 K variations in temperature over a number of Cassini
flybys between 2004 and 2010, and further evidence of wave activity on
large (150--450 km) vertical scales \citep{muller_06,
  snowden_13}. These perturbations in temperature were found to be
much larger in the upper atmosphere than in the lower mesosphere and
stratosphere \citep{lorenz_14}. Additionally,
INMS measurements revealed warmer upper atmospheric temperatures on
Titan's nightside, and while it was within Saturn's plasma sheet;
cooler temperatures were found on both the dayside and when Titan was
in Saturn's magnetospheric lobe, though no strong correlation
was found between temperature and latitude, longitude, solar zenith angle or local solar time \citep{cui_09, westlake_11, snowden_13}. UV
occultations sounding Titan's middle atmosphere measured a mesopause
at $\sim650$ km \citep{shemansky_05, liang_07},
and temperature inversions between 350--500 km attributed to Titan's
seasonally variable detached haze layer \citep{sicardy_06, lavvas_09},
which was also detected with ISS
and HASI \citep{porco_05, fulchignoni_05, west_11}.

As a strong contributor to both cooling Titan's upper atmosphere and producing
further, complex organic species, HCN provides substantial insight into
Titan's innately linked chemical and dynamical systems. HCN was first
detected on Titan by Voyager 1 \citep{hanel_81}, and has been
routinely observed in the millimeter regime
(see, for example: \citealp{paubert_87, hidayat_97, marten_02,
  courtin_11, rengel_22}). Following
observations of CH$_4$ fluorescence in Titan's mesosphere
--- resulting in additional measurement of Titan's
mesopause near 600 km \citep{kim_00} --- HCN
fluorescent emission was
observed with the NIRSPEC instrument on Keck II by \citet{geballe_03} and
subsequently re-analyzed by \citet{yelle_03} and \citet{kim_05} to produce
vertical profiles in Titan's mesosphere and thermosphere. These
measurements, and those from the Cassini VIMS, UVIS, and
INMS instruments, revealed Titan's HCN mole fraction to range between
$\sim3\times10^{-4}$--$3\times10^{-3}$ at 1000 km, along with
evidence for diurnal variability \citep{shemansky_05,
  magee_09, koskinen_11,
  adriani_11, vinatier_15, cui_16}. In the stratosphere and mesosphere, the variability of Titan's HCN abundance has been studied
using Cassini/CIRS, revealing
dramatic enhancements at high latitudes larger than an order of
magnitude throughout Titan's seasonal cycle \citep{teanby_07a,
  vinatier_15, teanby_19}. Further discrepancies exist between these
observations and predictions from photochemical models
(e.g. \citealp{loison_15, willacy_16, vuitton_19}), raising additional
questions pertaining to the role of HCN in Titan's chemical and radiative balance. 

After the end of the Cassini/Huygens mission, ground-based
observations continue to provide insight into Titan's complex
atmospheric composition and dynamics on short-term and seasonal
timescales. In particular, the Atacama Large Millimeter/submillimeter
Array (ALMA) has proven to be a powerful facility to study planetary
atmospheres, and allows for the study of Titan's stratosphere through
thermosphere by way of strong rotational emission lines from a variety
of molecular species. Previous ALMA observations from 2012 to 2015 allowed for the study
of temperature and HCN abundance in Titan's stratosphere during its
northern spring \citep{serigano_16, molter_16, thelen_18, thelen_19a},
while spectral Doppler shift measurements revealed
short-term variability in Titan's wind field in the middle and upper
atmosphere (\citealp{lellouch_19} --- hereafter, \citetalias{lellouch_19}; \citealp{cordiner_20}). Additionally, through the
combination of carbon monoxide (CO) and HCN emission lines, \citetalias{lellouch_19} retrieved Titan's vertical temperature and HCN
abundance profiles near the equator during 2016, showing a distinct mesopause at $\sim800$ km.

Here, we analyze ALMA observations of Titan from 2012, 2014, 2015, and
2017 ($L_S\sim34$--$89^{\circ}$) to determine the thermal structure and HCN abundance throughout
Titan's upper atmosphere. These results are compared to previous
measurements from the Cassini and Voyager missions, ground-based
observations, and model predictions of Titan's atmosphere. We describe the observations and data
reduction in Section \ref{sec:obs}, and the radiative transfer
modeling process in Section \ref{sec:rt}. The results, comparisons to
previous measurements and models, and discussion are presented in
Section \ref{sec:d}, followed with concluding remarks in Section \ref{sec:conc}.

\section{Observations} \label{sec:obs}
Data from the ALMA main-array, comprised of up to 50 12-m antenna dishes,
were obtained from the ALMA Science
Archive\footnote{https:$//$almascience.nrao.edu$/$aq$/$} covering a
period of 2012 through 2017. Observations of Titan were chosen such that the ALMA
spatial resolution, often denoted by the full-width at half-maximum
(FWHM) of the point-spread function (PSF) --- i.e. the synthesized beam
--- was \textless0.5$''$ (roughly half the angular size Titan's solid
body plus extended atmosphere subtends on the sky); the spectral resolution was
high enough to resolve the $\sim10$ MHz wide HCN line core, enabling measurement of
Titan's upper atmospheric temperature profile \citepalias{lellouch_19}; the
same rotational transition of HCN was measured across multiple
observations so as to directly compare the derived vertical HCN and temperature
profiles --- in this case, the HCN $J=4-3$ transition at 354.505 GHz
was used. The resulting parameters for selected archival observations with
the qualities specified above are presented in Table \ref{tab:obs},
including a mix of dedicated Titan observations from 2012, 2016, and
2017, and short observations from 2014 and 2015 where Titan was used
as a flux calibration source. As the 2016 data were previously
analyzed by \citetalias{lellouch_19}, the observational parameters and
spectra are only shown here for context and comparison.

\begin{deluxetable}{lllllllll}[b]
    \tablecaption{ALMA Observational Parameters}
    \tablecolumns{9}
    \tablehead{Project & Obs. & Int. & Spec. & Chan.
      & 
      Spatial & Pos. & Sub-Sol. & $L_S$$^{b}$\\
       Code & Date & Time & Res. & Width & Res.$^{a}$ & Ang. 
       & Lat. & \\
       & & (s) & (kHz) & (kHz) & ($''$) & ($^{\circ}$) & ($^{\circ}$)
       & ($^{\circ}$) \\ [-2.5ex]}
    \startdata
    2011.0.00727.S & 2012 Jun. 15 & 1572.6 & \phantom{1}122 & \phantom{11}61 &
0.470$\times$0.365 & \phantom{-}52.60 & 12.64 & 34.34 \\ 
2012.1.00635.S & 2014 Apr. 25 & \phantom{1}312.5 & \phantom{1}976 &
\phantom{1}488 &
0.457$\times$0.380 & -81.53 & 22.10 & 55.68 \\ 
2013.1.00446.S & 2015 Jun. 12 & \phantom{1}151.3 & \phantom{1}282 &
\phantom{1}244 & 
0.361$\times$0.304 & -67.06 & 24.28 & 68.41 \\ 
2015.1.01023.S$^{c}$ & 2016 Aug. 19 & 2925.9 & 1938$^{d}$ & 1953 & 0.265$\times$0.144 & \phantom{-}52.13 & 26.07 & 81.59 \\ 
2016.A.00014.S & 2017 May 08 & 1080.0 & \phantom{1}976$^{d}$ & \phantom{1}488 &
0.231$\times$0.185 & -75.85 & 26.41 & 89.49 \\ 

    \enddata
    \footnotesize
    \tablecomments{$^{a}$Spatial resolution corresponds to the FWHM of the ALMA
      PSF. $^{b}$Refers to the solar longitude (measured from the
      vernal equinox) 
      of Saturn. $^{c}$Previously analyzed by
      \citetalias{lellouch_19}, shown here for reference. $^{d}$Higher
      spectral resolution windows centered just on the HCN line core
      were also included in these observations for the measurement of winds.}
    \label{tab:obs}
  \end{deluxetable}

The ALMA observations were reduced and calibrated in the Common
Astronomy Software Application (CASA) package
\citep{jaeger_08} using the Joint ALMA
Observatory (JAO) pipeline scripts provided with each
observation. Typical pipeline calibration procedures included the
correction of complex visibility gain and phase measurements, and
amplitude calibration using strong (sub)millimeter continuum sources,
such as quasars or Solar System moons (including Titan
itself). For observations from 2014 and 2015, where Titan was used as
a flux calibration target, JAO scripts were modified to remove commands
flagging spectral windows containing HCN emission lines, which are
often removed during calibration procedures due to their strong, broad line
wings. Calibrated visibilities were corrected to Titan's rest
velocity frame using ephemeris data from the JPL Horizons
System\footnote{https:$//$ssd.jpl.nasa.gov$/$horizons$/$app.html$\#/$}
and deconvolved to remove interferometric artifacts
using the CASA \texttt{tclean} task. During imaging, pixel sizes were
set to $\frac{1}{4}$ to $\frac{1}{5}$ of the ALMA synthesized beam size. The H{\"o}gbom clean
algorithm \citep{hogbom_74} was used, and additional weighting to long antenna baselines
was applied using the Briggs weighting scheme with a robust parameter
of 0.5 \citep{briggs_95a}. The data were cleaned to a flux threshold of $2\times$ the RMS noise.  

\begin{figure}
  \centering
  \includegraphics[scale=1.0]{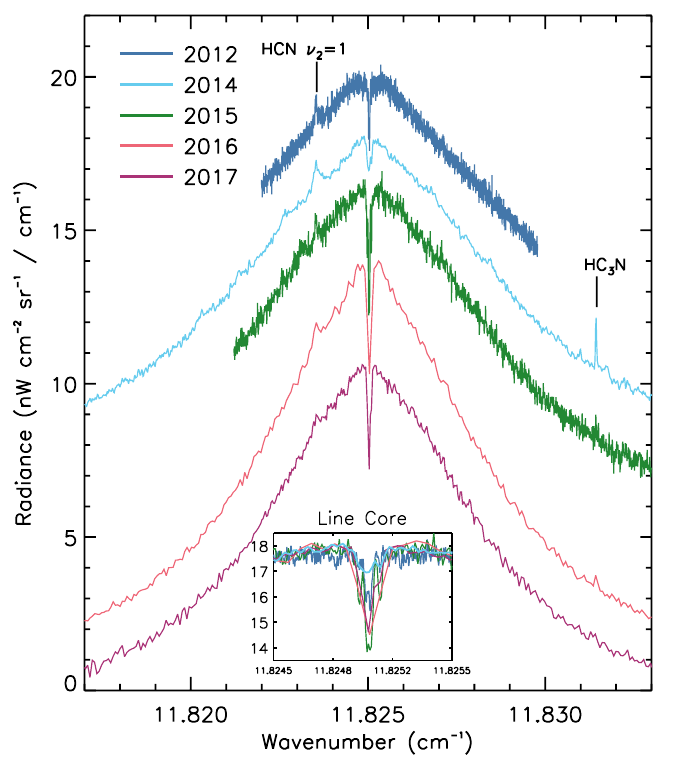}
  \caption{Titan HCN ($J=4-3$) nadir spectra from ALMA observations in 2012 (blue),
    2014 (cyan), 2015 (green), 2016 (red), and 2017 (purple). Spectra
    are vertically offset above or below the 2014 spectrum by subsequent
      multiples of 2 nW cm$^{-2}$
      sr$^{-1}$ / cm$^{-1}$ for clarity, and are plotted in radiance units to
    account for the change in beam size between
    observations. The inset shows comparisons of absorption in the HCN line
    core, which sounds Titan's mesopause, and the spectra have been
    shifted to compare absorption depths.}
  \label{fig:spec_f}
\end{figure}

Individual spectra were extracted from pixels at Titan's nadir and western limb, and converted to radiance units for use during radiative
transfer modeling using the procedures described in \citet{thelen_18,
  thelen_19a}. Nadir spectra of Titan's strong, rotational HCN
emission lines are particularly sensitive to the mesospheric
temperature structure \citepalias{lellouch_19}; comparisons of HCN nadir
spectra are shown in Figure \ref{fig:spec_f}, with the inset showing
differences in the line core depth between each
observation. Simultaneously modeling both limb and nadir spectra
allows for the retrieval of Titan's vertical thermal profile while
mitigating degeneracy imposed by variability in the vertical HCN
volume mixing ratio (VMR) profile \citepalias{lellouch_19}. So as to
ensure the HCN line core was fixed at the rest frequency, small spectral line
shifts were added to extracted limb spectra (on the order of 1--2$\times$
the ALMA channel spacing, Table \ref{tab:obs}) to account for Doppler
shifts induced by upper atmospheric winds found in high resolution observations of various molecular
species (\citetalias{lellouch_19}; \citealp{cordiner_20}).

\section{Radiative Transfer Modeling} \label{sec:rt}
\begin{figure}
  \centering
  \includegraphics[scale=0.8]{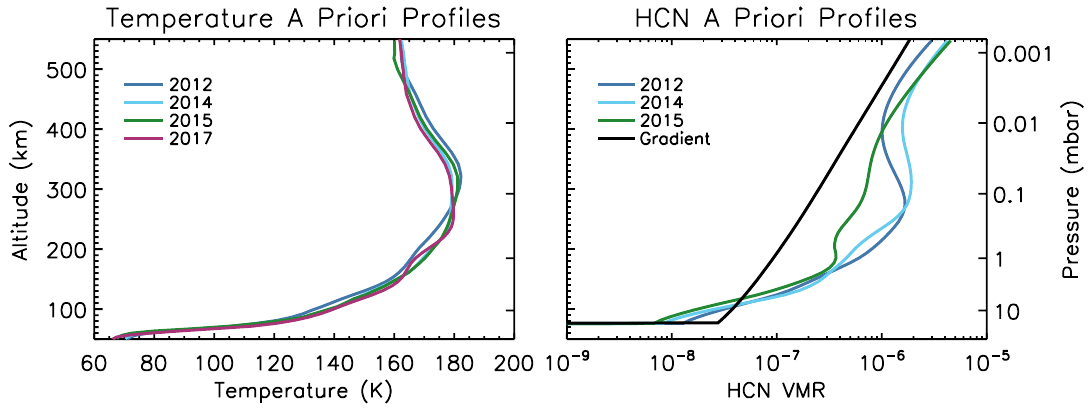}
  \caption{\textit{A priori} vertical profiles used for radiative transfer
    modeling. (Left) Temperature
  \textit{a priori} profiles from \citet{thelen_18, thelen_agu_19} retrieved from
  ALMA CO spectra (100--550 km) and Cassini Radio Science data (\textless 100 km) in 2012 (blue), 2014 (cyan), 2015 (green), and 2017 (purple). (Right) HCN VMR profiles retrieved
    from ALMA H$^{13}$CN and HC$^{15}$N spectra in 2012, 2014, and
    2015 from \citet{thelen_19a}. A simple vertical gradient (black) was used to test the sensitivity
  of retrievals to vertical structure. }
  \label{fig:ap}
\end{figure}

As rotational HCN emission lines are pressure-broadened by
  Titan's atmosphere and thus vary in line-shape due to changes in
  atmospheric state with altitude (and, as a result of the ALMA beam
  size, latitude), radiative transfer models were calculated
  independently for spectra from each observation to determine both the continuous
  vertical HCN VMR and temperature profiles from the top of the
  stratopause to the lower thermosphere.
Modeling was performed using the Non-linear optimal
Estimator for MultivariatE spectral analySIS (NEMESIS) radiative
transfer code developed by \citet{irwin_08}. Synthetic HCN spectra were calculated
implicitly using NEMESIS in line-by-line mode, using spectral line and
partition function data from the Cologne Database for Molecular Spectroscopy\footnote{https:$//$cdms.astro.uni-koeln.de}
\citep{muller_01} and HITRAN catalogue\footnote{https:$//$hitran.org}
\citep{gordon_22}. Line broadening and temperature dependence
coefficients, collisionally-induced absorption pairs, and initial gas
profiles for N$_2$,
CH$_4$, and H$_2$, were taken from \citet{serigano_16},
\citet{molter_16}, \citet{thelen_18}, and references
therein. \textit{A priori} vertical profiles for CO, HCN, and HC$_3$N
were used from ALMA
observations of Titan in 2012, 2014 and 2015 presented in previous analyses \citep{molter_16, serigano_16, thelen_18,
  thelen_19a, thelen_agu_19}. The temperature and HCN
\textit{a priori} vertical profiles used here are shown in Figure \ref{fig:ap}.

Models of Titan's (sub)millimeter thermal emission were initialized
using temperature profiles from Cassini Radio Science observations,
sensitive to Titan's troposphere and stratosphere, at
altitudes \textless 100 km \citep{schinder_12,
  schinder_20}. Spectral windows adjacent to the HCN $J=4-3$ emission
line were then used to determine offsets between the data and model
continuum to derive a multiplicative corrective scaling factor (if needed), as
described in \citet{thelen_18}; these corrective factors were on order
0.90--1.10, a typical uncertainty for flux density calibrations of ALMA with quasars
and Solar System objects \citep{fomalont_14}.

\begin{figure}
  \centering
  \includegraphics[scale=0.8]{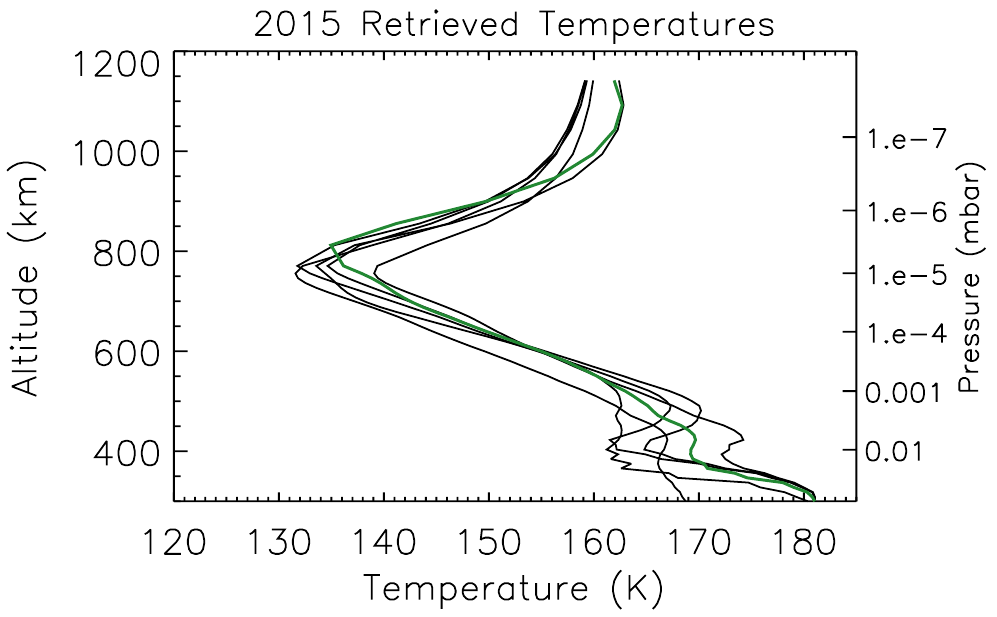}
  \caption{Comparison of retrieved temperature profiles from 2015
    after using perturbations of the \textit{a priori} temperature and HCN
    profiles shown in Fig. \ref{fig:ap} to illustrate 
    the stability of the retrieved mesopause and lower mesospheric
    inversion near 400 km ($\sim0.01$ mbar). The best fit vertical
    profile is shown in green.}
  \label{fig:2015_temps}
\end{figure}

As previous studies retrieved Titan's vertical temperature (using CO) and
HCN profiles throughout Titan's stratosphere \citep{serigano_16, molter_16,
  thelen_18, thelen_19a}, we parameterized NEMESIS models of HCN
spectra such that the vertical temperature and HCN abundance were allowed to vary
continuously above the stratopause ($\sim250$--350 km) in a grid of
108 atmospheric layers. Layers were separated vertically by 3--40 km,
increasing with altitude. Simple \textit{a priori} vertical profiles above the stratosphere 
were used so as to avoid imparting artificial vertical structure on the
retrieved profiles. These include an isothermal profile at 160 K at
altitudes \textgreater600 km, and a
simple vertical HCN gradient in log(VMR) from its condensation altitude at $\sim80$ km
\citep{marten_02} up to a VMR = $2\times10^{-4}$ at 1100 km
\citep{vuitton_07}. Subsequent model runs incorporated perturbations to the HCN profile by an order of
magnitude in either direction, or used the photochemical model profile of
\citet{loison_15}, to test the sensitivity of the retrievals to
vertical structure (\citealp{molter_16}; \citetalias{lellouch_19}). Unique vertical features, such as a small temperature inversion
around 400 km (Figure \ref{fig:2015_temps}), were considered valid if
they manifested in the majority of retrievals. The correlation length for temperature and HCN
profiles was set to 1.5 and 3.0 scale heights, respectively, so as to
prevent unphysical vertical oscillations in the retrieved HCN profiles
\citep{thelen_18, molter_16}. 1-$\sigma$ errors on the vertical thermal profile
were initialized at 1 K above the stratopause and raised to 5--10 K in the upper
atmosphere, while the HCN \textit{a priori} errors were set at 100$\%$ of the HCN VMR. 

\begin{figure}
  \centering
  \includegraphics[scale=0.7]{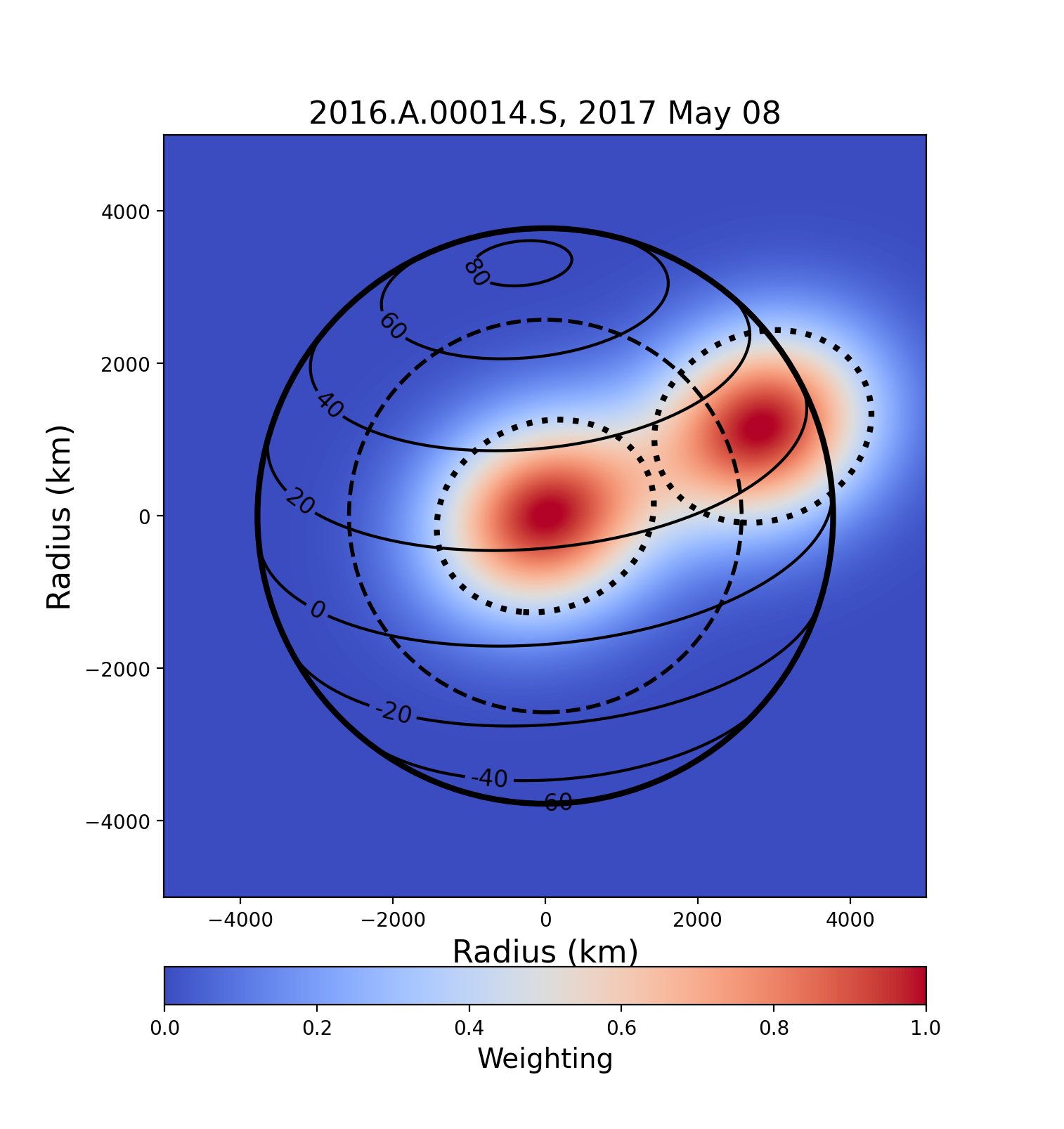}
  \caption{Weighting function distribution for nadir and limb viewing
    geometries for the 2017 observation. Beam locations were placed
    such that the top-of-atmosphere latitudes matched for both
    locations for each observation, taking into account Titan's tilt. Weighting distributions are
    not coadded, as plotted, but calculated individually for each extraction
    region. Titan's solid body radius
    (2575 km; dashed circle), top of atmosphere (1200 km above the surface; solid circle), and
    top-of-atmosphere latitudes are shown. The FWHM of the ALMA beam
    is denoted by the dashed ellipses.}
  \label{fig:bms}
\end{figure}

\begin{figure}
  \centering
  \includegraphics[scale=0.43]{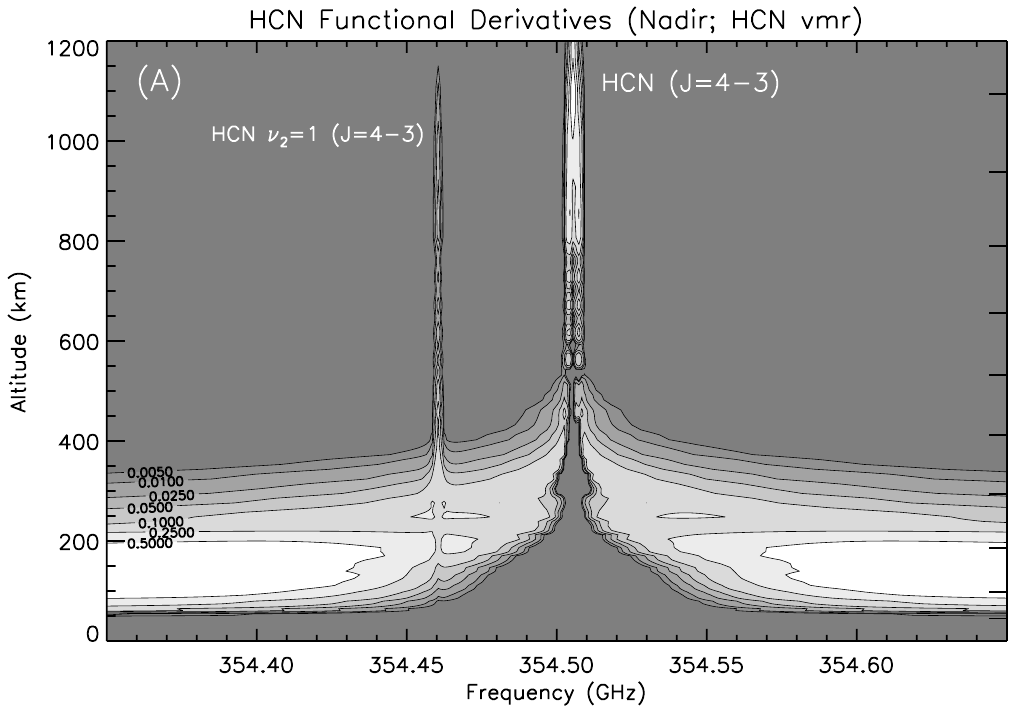}
  \includegraphics[scale=0.43]{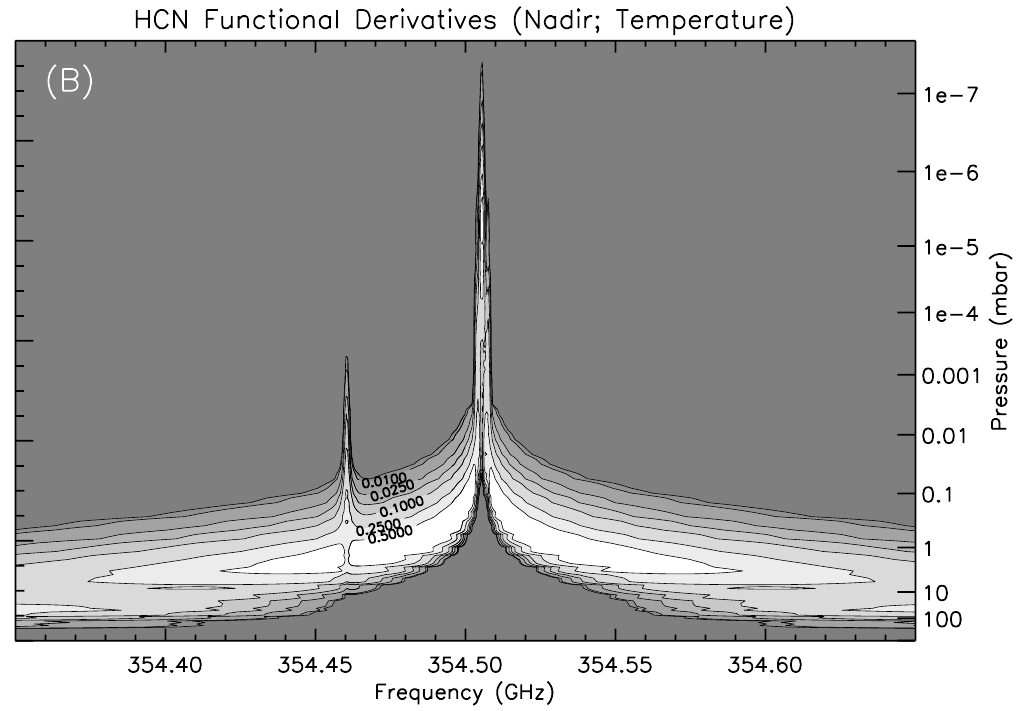}
  \includegraphics[scale=0.43]{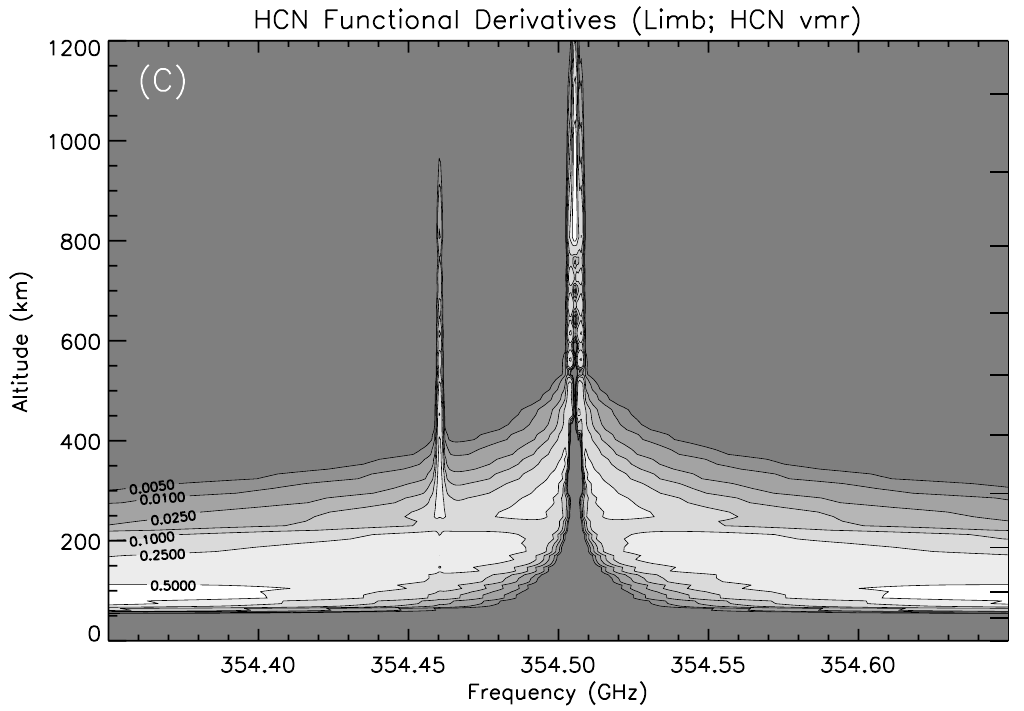}
  \includegraphics[scale=0.43]{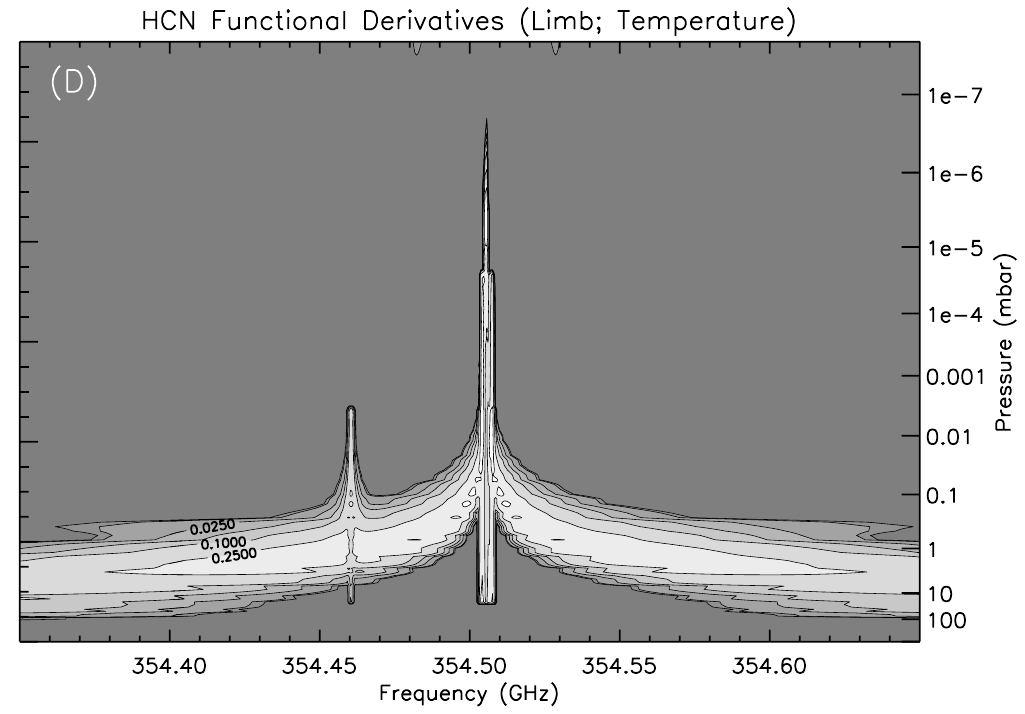}
  \caption{Plots of normalized functional derivatives calculated by
    NEMESIS for HCN ($J=4-3$) spectra as a function of frequency and
    altitude with respect to HCN VMR (left
    column) and temperature (right column) in nadir and limb viewing
    geometries (top row and bottom row, respectively). Contours 
    increase with relative contribution.}
  \label{fig:cf}
\end{figure}

As in \citetalias{lellouch_19}, the HCN abundance and temperature were
retrieved simultaneously by fitting both nadir and limb spectra at
once. While previous observations have shown Titan's thermospheric
temperature and HCN abundance to vary with longitude
\citep{snowden_13, cui_16}, we found the
retrieved profiles from East and West limbs to fall within the
corresponding retrieval errors. To accurately model emission from the ALMA beam shape, 20
emission angles were used for each viewing geometry, as described in
\citet{thelen_18}\footnote{The ALMA spectrum extraction and emission
  angle weighting code may be found at:
  https://data.mendeley.com/datasets/szbcb44s43/1
  \citep{thelen_17_br}.}. A representative set of emission angle weights are shown in
Figure \ref{fig:bms} for the observation from 2017 May 08 (ALMA Project
Code 2016.A.00014.S). A
smaller range of data points was modeled near the HCN line center at
the native ALMA spectral resolution, allowing for higher weighting of the line core during the
$\chi^2$-minimization of iterative spectral
inversions, thereby optimizing the sensitivity of the retrievals to
the relevant atmospheric pressures. Derivatives
calculated for the HCN $J=4-3$ limb and nadir spectral radiances as a function of both HCN VMR and
temperature are shown in Figure \ref{fig:cf} (see Section 2.1 of
\citet{irwin_08} for a complete description of functional derivatives
in the NEMESIS code). These functional
derivatives illustrate the
sensitivity of the HCN line core to Titan's temperature profile up to
$\sim1000$ km, and the complementary nature of limb and nadir spectral
sensitivity to HCN abundance that helps to break degeneracy in HCN and temperature
retrievals \citepalias{lellouch_19}. The resulting best fit
limb and nadir spectra are shown in Figure \ref{fig:spec_c}; see the
Supplementary Materials of \citetalias{lellouch_19} for the analysis of data from 2016. 

\begin{figure}
  \centering
  \includegraphics[scale=1.0]{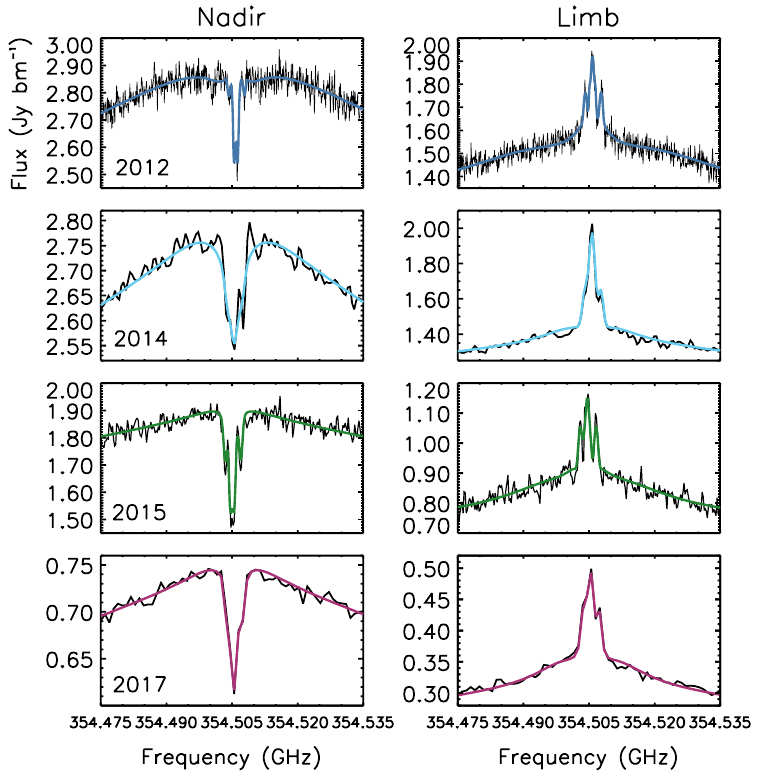}
  \caption{ALMA spectra (black) extracted from nadir (left column) and limb
    (right column) pixels compared with the best fit NEMESIS
    models (colored lines) for 2012 (row 1), 2014 (row 2), 2015 (row
    3) and 2017 (row 4).}
  \label{fig:spec_c}
\end{figure}

\section{Results and Discussion} \label{sec:d}
\begin{figure}
  \centering
  \includegraphics[scale=0.8]{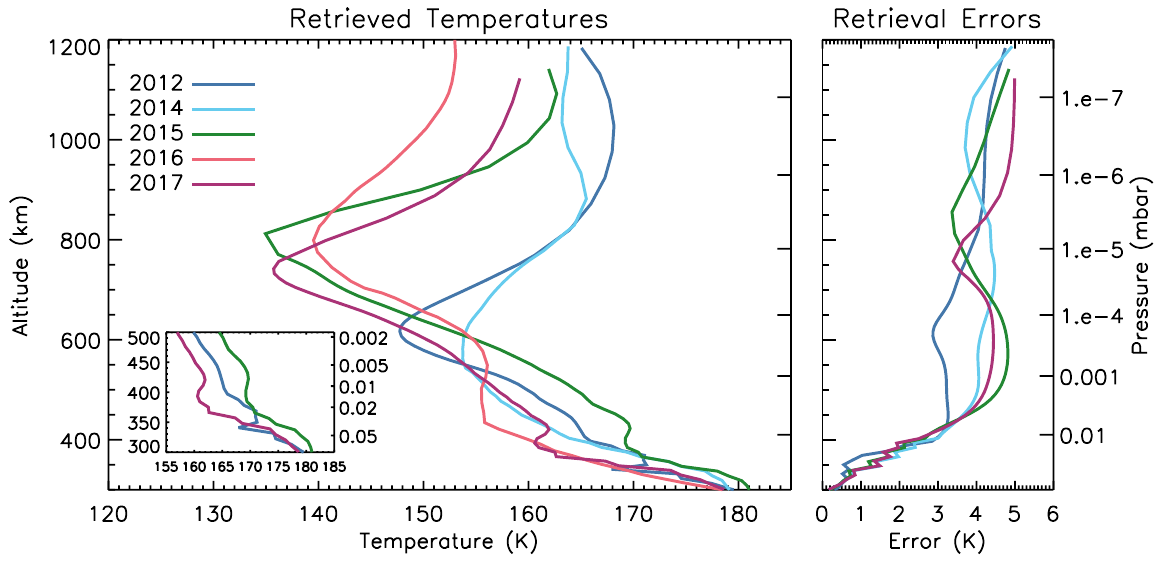}
  \caption{(Left) Retrieved vertical temperature profiles corresponding to
    the NEMESIS best fit models in Figure \ref{fig:spec_c}. The retrieved temperature
    profile from \citetalias{lellouch_19} is shown in red, for
    comparison. (Inset) Comparison of the retrieved 2012, 2015, and
    2017 profiles near the location of Titan's detached haze layer. (Right) Errors
    on the retrieved 2012, 2014, 2015, and 2017 models. The pressure
    scale on the right y-axis is approximate.}
  \label{fig:temps}
\end{figure}

\begin{figure}[h]
  \centering
  \includegraphics[scale=0.8]{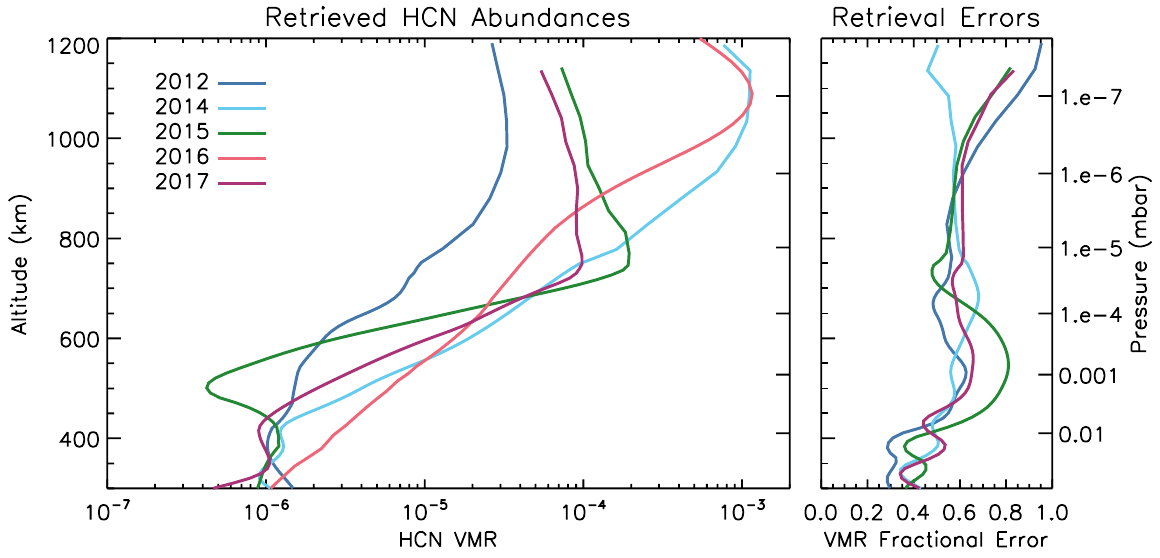}
  \caption{(Left) Vertical HCN VMR profiles corresponding to
  the NEMESIS best fit models in Figure \ref{fig:spec_c}, retrieved
  simultaneously with the temperature profiles in
  Figure \ref{fig:temps}. The retrieved HCN profile from
  \citetalias{lellouch_19} is shown in red, for comparison. (Right)
  Fractional errors on the retrieved 2012, 2014, 2015, and 2017 VMR
  profiles. The pressure
    scale on the right y-axis is approximate.}
  \label{fig:hcn}
\end{figure}

The retrieved vertical profiles corresponding to the best fit spectral
models presented in Figure \ref{fig:spec_c} are
shown in Figure \ref{fig:temps} and \ref{fig:hcn}, accompanied by the
retrieved temperature and HCN profiles from 2016 adapted from
\citetalias{lellouch_19}. As with previous observations of Titan's
mesosphere and thermosphere, substantial variability is present
between the four profiles derived here and that of
\citetalias{lellouch_19} above $\sim400$ km (0.01 mbar),
reaching maximum differences of 10--30 K. The differences across our
retrieved temperatures are much larger than those found for the
stratosphere during the same period with similar ALMA observations
(generally $\sim5$ K from Earth year to year; Figure \ref{fig:ap}, left; \citealp{thelen_18}). A distinct mesopause
was found for all 4 years analyzed here and in the profile derived
from 2016 data by \citetalias{lellouch_19}, as indicated by the
presence of absorption in the HCN line core (Figure \ref{fig:spec_f}, inset) that is distinctly
sensitive to the pressure, temperature, and HCN abundance of the atmosphere \textgreater600 km (Figure \ref{fig:cf}). The location and temperature of
these relative minima are largely different from one observation to
the next. This is
also clearly visible in the raw spectra as differences in relative depth and shape of the HCN
line cores. The warmest and lowest temperature
minimum, $153.7 \pm 4.0$ K at $563 \pm 10$ km
($\sim$3.8$\times10^{-4}$ mbar), was found in 2014 ($L_S\sim56^{\circ}$). The coldest and highest
mesopause, $134.9 \pm 3.4$ K at $812 \pm 42$ km ($\sim$3.3$\times10^{-6}$
mbar), was found in 2015 ($L_S\sim68^{\circ}$), just under 26 Titan days (413 Earth days) later. The 2012 and 2017
measurements, and those found by \citetalias{lellouch_19}, are
bracketed by these extremes, resulting in a distinct and rapid
separation between the 2012, 2014 ($L_S\sim34$--$56^{\circ}$)
observations and those after 2015 ($L_S\ge68^{\circ}$). The upper
atmospheric temperatures vary between 153 K (2016;
\citetalias{lellouch_19}) and $165 \pm 4.4$ K
(2014) at 1200 km ($2.0\times10^{-8}$ mbar), while the largest temperature difference in the
lower mesosphere is found at $\sim$450 km ($4.6\times10^{-4}$
mbar) between 2015 and 2016, where the temperature varies between $155 \pm 3.8$ K 
and $169 \pm 3.0$ K. \citetalias{lellouch_19} note a stable,
quasi-isothermal region in their retrieved profiles from
$\sim450$--600 km, which was also evident in the \textit{a
  priori} profiles they employed based on data from Cassini/CIRS. Similarly, a small temperature inversion was found to be stable in our
2012, 2015, and 2017 results, as shown in the inset in
Figure \ref{fig:temps}. These inversions may be attributed to Titan's detached or
transient haze
layers, as discussed further in Section \ref{sec:dhl}. 

\begin{figure}
  \centering
  \includegraphics[scale=0.83]{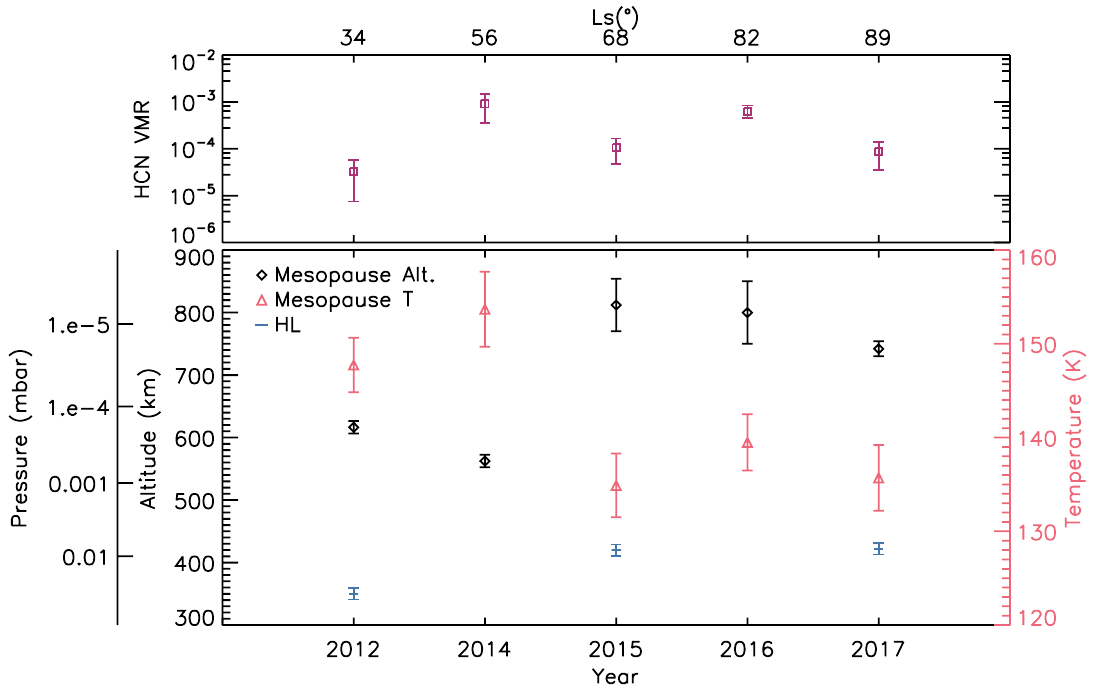}
  \caption{(Top) HCN VMR at 1000 km as a function of time. (Bottom)
    Mesopause (black diamonds) and haze layer (blue lines) altitudes
    (left y-axis) and mesopause temperatures (red triangles,
    right y-axis) as a function of time. The secondary pressure axis is
  approximate.}
  \label{fig:tc}
  \end{figure}

The retrieved vertical HCN profiles, as shown in Figure \ref{fig:hcn},
show similarly large variability between the 4 observations analyzed
here. In particular, the 2012 spectrum required substantially lower
HCN abundance throughout the atmosphere above $\sim600$ km
($2.5\times10^{-4}$ mbar), resulting in an order of magnitude less HCN
in the lower atmosphere than the other retrievals and $\sim30\times$
less than the 2014 and 2016 \citepalias{lellouch_19} profiles near 1100 km
($7.6\times10^{-8}$ mbar). Attempts to fit the spectrum from 2012
using the retrieved HCN profiles from 2014 as an \textit{a priori}
resulted in large reduced $\chi^2$ values for spectral fits, and required significant or
physically unlikely deviations from the input profiles. This suggests that the atmosphere in Titan's low-northern
latitudes during 2012 was depleted of HCN above the mesopause
(occurring near this location in 2012) by a factor of 2--30. Similarly, the 2015 and 2017
retrievals require roughly an order of magnitude less HCN above
$\sim800$ km ($7.4\times10^{-6}$ mbar) than those of 2014 and 2016
\citepalias{lellouch_19}. The retrieved HCN VMR values at 1000 km, along with the mesopause
altitudes and temperatures, are plotted as a function of time in
Figure \ref{fig:tc}. While the distinct transition between
  the 2014 and 2015 mesopause locations and magnitudes is evident
  here, a correlation with the HCN abundance is not apparent. 

\subsection{Comparisons to Previous Studies} \label{sec:comp}
\begin{figure}
  \centering
  \includegraphics[scale=0.75]{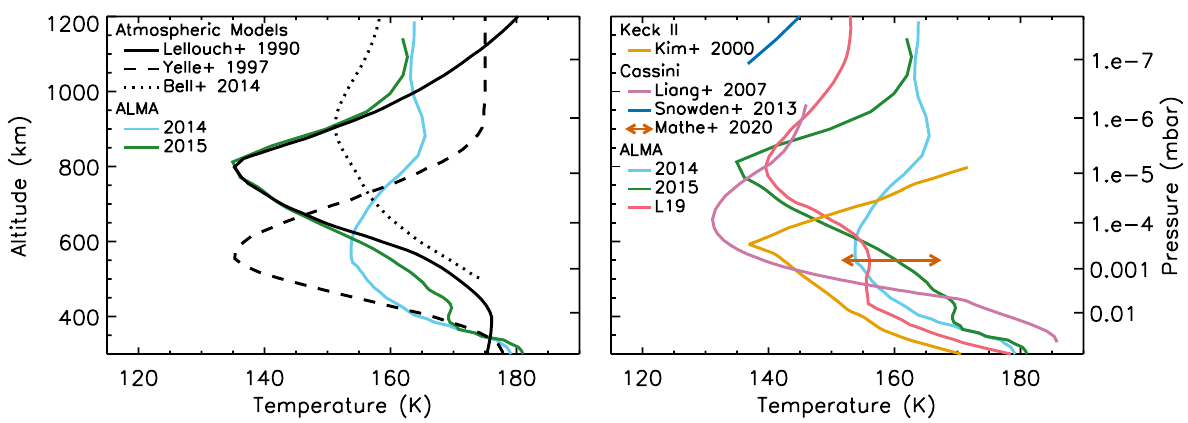}
  \caption{Comparison of the retrieved 2014 ($L_S\sim56^{\circ}$; cyan) and 2015
    ($L_S\sim68^{\circ}$; green) temperature profiles to: (Left) atmospheric model
    predictions from \citet{lellouch_90}, \citet{yelle_97}, and
    \citet{bell_14} (their Model C); (Right) ground-based observations
    (\citealp{kim_00}; \citetalias{lellouch_19}), and measurements
    from the Cassini UVIS \citep{liang_07},
    INMS \citep{snowden_13}, and CIRS (\citealp{mathe_20}, shown as a
    range of equatorial temperatures measured from 2010 to 2017) instruments. The pressure
    scale on the right y-axis is approximate.}
  \label{fig:lit_t}
\end{figure}

Our results are compared to previous measurements and predictions of Titan's thermal
profile in Figure \ref{fig:lit_t}. Titan's mesospheric
temperature profiles appear to transition between two states, with a
mesopause at $\sim$600 or 800 km, similar to previous
measurements by \citet{kim_00} and \citet{liang_07} (revised from
\citealp{shemansky_05}) in 1999 and 2004, respectively. Our profiles
also agree with the altitudes of the mesopause predicted by the
atmospheric models of
\citet{lellouch_90} and \citet{yelle_97}, though the magnitude of the
mesopause temperature is only matched by the \citet{lellouch_90} model
in 2015; our 2014 profile is significantly warmer ($\sim20$ K) than
the results of \citet{yelle_97}, despite the similar location of the
inversion. These comparisons may reflect the differences in the
incorporated physics and inclusion of HCN cooling between the two
models \citep{yelle_91}, while the discrepancy with the model of
\citet{yelle_97} may be due to assumptions of or seasonal variations in aerosol heating, 
CH$_4$ or C$_2$H$_6$ abundances, which affect the temperature profile
more strongly at $\sim600$ km \citep{yelle_91}. Our retrieved mesopause temperatures are generally warmer than
previous observations from earlier in Titan's seasonal cycle, though they occur at similar
altitudes, with the exception of the profile derived from HASI
measurements during the Huygens descent, where a temperature minimum
of 152 K was found at $\sim$490 km. We find warmer upper atmospheric
temperatures than the average profile measured by Cassini/INMS,
though within one standard deviation (125--165 K
at 1200 km) observed over a large number of flybys
\citep{snowden_13}. The retrieved profiles are generally cooler than
the models of \citet{bell_10} and \citet{bell_14}, and do not
exhibit a mesopause at $\sim900$ km, as seen there; however,
the temperatures above those altitudes are generally in good
agreement. Our retrieved temperature profiles tend to relax
toward the \textit{a priori} value of 160 K above $\sim1100$ km, where
the spectra are less sensitive to deviations in the temperature
profile (Figure \ref{fig:cf}), consistent with
the thermospheric model of \citet{muller_08} at 1000 km for
low-northern latitudes ($164 \pm 6$ K at 20$^{\circ}$ N). The profiles derived from ALMA observations
between 2012 and 2017 are consistent with the range of values found through
contemporaneous observations with the 
Cassini/CIRS instrument in the lower mesosphere (see \citealp{mathe_20}, as depicted by the
range of temperatures in orange in Figure \ref{fig:lit_t}, and \citealp{vinatier_20}). Compared with the stratosphere, which varies
significantly at high latitudes and on seasonal timescales
\citep{sylvestre_20, sharkey_21}, the derived mesospheric temperature
profiles from ALMA observations present evidence for 
larger magnitude thermal variability within low-latitude regions on
shorter timescales.

\begin{figure}
  \centering
  \includegraphics[scale=0.75]{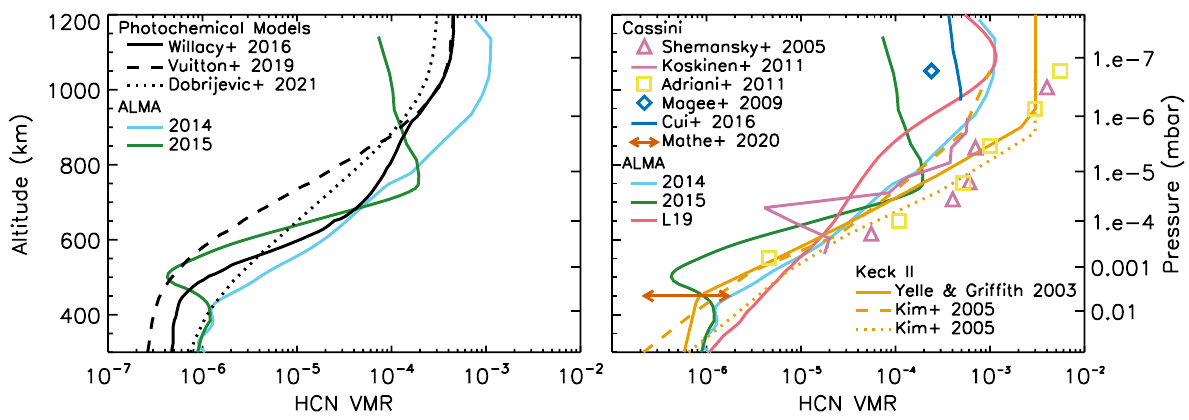}
  \caption{Comparison of the retrieved 2014 ($L_S\sim56^{\circ}$; cyan) and
    2015 ($L_S\sim68^{\circ}$; green) HCN vertical profiles to: (Left) photochemical
    model results \citep{willacy_16, vuitton_19, dobrijevic_21};
    (Right) ground-based observations (\citealp{yelle_03, kim_05}; \citetalias{lellouch_19}),
    Cassini UVIS \citep{shemansky_05, koskinen_11}, VIMS
    \citep{adriani_11}, INMS \citep{magee_09, cui_16}, and CIRS
    (\citealp{mathe_20}, as in Fig. \ref{fig:lit_t}) measurements. The pressure
    scale on the right y-axis is approximate.}
  \label{fig:lit_h}
\end{figure}

Previous measurements of Titan's upper
atmospheric HCN by ground-based observations and Cassini are compared
to our retrieved 2014 and 2015 ($L_S\sim56$ and $68^{\circ}$) profiles in
Figure \ref{fig:lit_h}. Results from Cassini UVIS and
VIMS observations \citep{shemansky_05, koskinen_11, adriani_11}, and those from the NIRSPEC
instrument on Keck II \citep{yelle_03, kim_05}, show generally higher
HCN abundances in the upper atmosphere than found with the Cassini
INMS instrument \citep{magee_09, cui_16} and predicted by
photochemical models \citep{willacy_16, vuitton_19,
  dobrijevic_21}. While our 2014 ($L_S\sim56^{\circ}$) retrieval and
that of \citetalias{lellouch_19} in 2016 ($L_S\sim82^{\circ}$)
fall somewhere in the middle of these prior observations, the 2012,
2015, and 2017 results deviate further still above $\sim900$ km,
extending the range of measured HCN in Titan's upper atmosphere from
$\sim3\times10^{-5}$ to $3\times10^{-3}$, with relative maxima
occurring at altitudes $\sim950$--1100 km. While 
atmospheric nitriles (e.g. HC$_3$N) have been found to vary with
latitude up to 2
orders of magnitude between Titan's equator and winter pole in the
stratosphere \citep{teanby_07a, vinatier_15}, we find that the
temporal HCN
variability is far more significant in the upper atmosphere throughout
its year compared
to contemporaneous retrievals in the stratosphere at low latitudes by
the Cassini CIRS instrument \citep{teanby_19, coustenis_20, mathe_20, vinatier_20}. 

\subsection{Interpretations} \label{sec:i}
\subsubsection{Upper Atmospheric Energetics and Dynamics} \label{sec:dy}
Compared to Titan's stratosphere, which
evolves on seasonal timescales \citep{flasar_81}, the 
upper mesosphere and thermosphere exhibit short-term variability due to the complex,
highly variable environment Titan encounters during its passage
through Saturn's non-uniform magnetosphere and gravitational potential
(due to Titan's orbital eccentricity $\sim0.03$), and its
changing insolation. Above 1000
km, we measure variations of 3--10 K between observations, likely
resulting from the combined effects of environmental factors such as changes in charged
particle precipitation (found to increase the temperature by $\sim7$
K by \citealp{snowden_14}) and Titan local time \citep{de_la_haye_08}. These changes typically affect
Titan's thermosphere on much shorter timescales than the time between observations
analyzed here
(on order of 1 Titan day; \citealp{bell_11, snowden_14}), so
while the effects of energy sources from the external environment are
present, they are difficult to quantify directly and compare from one observation to the next. 

No clear trend between the measured temperatures and Titan's
  orbital position with respect to the Saturnian magnetosphere is
apparent across the small sample size of data analyzed here. Titan's location during the 2014 and 2016 observations is
very similar --- to the west of Saturn, in the ram direction
with respect to the
Saturnian magnetosphere (see descriptions in \citealp{rymer_09} and \citealp{westlake_11}) --- though the thermal structure varies dramatically between
these two dates. Titan was observed completely outside of Saturn's
magnetosphere by Cassini during December 2013, roughly 8 Titan days
before the 2014 ALMA observation \citep{bertucci_15}. The effects of
the 2013 event, during which Titan was directly exposed to solar wind
particles, may be indirectly seen in the difference in the temperature
and HCN profiles between 2012 and 2014 observations. Similarly, this
time period also marked the maximum of solar cycle 24, which may increase
the EUV flux experienced by Titan compared to many of the Cassini
observations made during solar minimum \citep{yelle_14}.

The effects of HCN cooling are significant at altitudes
\textgreater700 km, as a variation in HCN abundance
by $\sim$an order of magnitude may cool the atmosphere by $\sim20$ K
\citep{yelle_91, bell_10, snowden_14}. This may help explain the warmest
thermospheric temperatures in the 2012 ($L_S\sim34^{\circ}$) profile, where HCN abundances
were found to be significantly less than the other
3 observations and the measurements of \citetalias{lellouch_19} ($L_S\sim82^{\circ}$). The thermal
profile of \citetalias{lellouch_19} is cooler than the profiles we
retrieved above $\sim950$ km and is accompanied by relatively high HCN
abundance in the thermosphere, while our profiles from 2015 and 2017
are warmer and require less HCN. The
temperature and HCN profiles from 2014 ($L_S\sim56^{\circ}$) remain puzzling, as the
thermospheric temperatures are much higher than those of
\citetalias{lellouch_19} and those predicted by the model of
\citet{snowden_14} for similar HCN volume mixing ratios. The large
change in HCN abundance from 2012 to 2014 ($L_S\sim34^{\circ}$ to $56^{\circ}$), and the seeming imbalance
between its predicted cooling on the thermosphere, may be the result of the
recent change in Titan's post-equinox stratospheric circulation,
disruption of polar vortices, or
other energetic events during the northern spring period
\citep{teanby_12, vinatier_15, teanby_17, coustenis_18,
  teanby_19, 
  coustenis_20, sharkey_20, vinatier_20,
  sharkey_21}.

While the changes between temperature profiles from $\sim400$--600 km
appear anti-correlated with changes in the HCN VMR
profiles, implying some cooling at these altitudes due to
increased HCN atmospheric content, this effect is less efficient at
these altitudes according to the model of \citet{yelle_91}, and not enough to explain 
the largest difference of $\sim15$ K between the 2015 and 2017
profiles. At these altitudes, vibrational emission
from C$_2$H$_6$ also becomes important, though the profiles retrieved
in the lower mesosphere with Cassini/CIRS do not show significant
dispersion at low latitudes during this time period
\citep{mathe_20}. Unfortunately, the balance between HCN,
C$_2$H$_2$, CH$_4$, and C$_2$H$_6$ cooling in the lower mesosphere is
difficult to quantify with observations by ALMA due to
the lack of observable transitions for these species in the (sub)millimeter.

\subsubsection{Detached Haze Layer} \label{sec:dhl}
Images of Titan from the Voyager and Cassini missions revealed the
presence of a vertically variable, detached haze layer \citep{rages_83,
  porco_05, west_11, west_18, seignovert_21}, later linked to the top of Titan's
stratospheric meridional circulation \citep{rannou_04, lebonnois_12, larson_15}
as its seasonal evolution cannot be explained by microphysical processes alone \citep{cours_11}. The efficient absorption of solar photons by
the dense haze in Titan's lower mesosphere was also found to manifest
as a sharp inversion in thermal profiles derived from ground-based and
Cassini/UVIS stellar occultations as well as in the Huygens/HASI data 
\citep{sicardy_99, fulchignoni_05, sicardy_06, lavvas_09,
  koskinen_11}. High cadence images near Titan's 2009 spring equinox by
Cassini/ISS showed the altitude of the detached haze layer descending
rapidly from $\sim500$ km to $\sim350$ km between 2008 and 2011
along with the collapse of the main haze \citep{west_11, seignovert_17, west_18, seignovert_21},
resulting in similar post-equinox altitudes to that observed by Voyager 2 after
Titan's spring equinox in 1980 (357 km at the equator; \citealp{rages_83}). The consistent haze
altitudes between 300--500 km, and their subsequent descent near
equinox, were corroborated by Cassini/UVIS and ground-based
occultations during both the 1980 and 2009 spring equinoxes
\citep{sicardy_99, sicardy_06, liang_07, koskinen_11} and through general circulation models of Titan's
atmosphere \citep{rannou_02, lebonnois_12, larson_15}. However, the
temporal and spatial nature of the occasionally multiple, transient
detached haze layers and plumes indicates a complex relationship between dynamics
and microphysics that has proven difficult to predict on short time
scales through current modeling efforts. 

In the ALMA temperature profiles presented here, we find similar
inversions in the 2012, 2015, and 2017 ($L_S\sim34, 68, 89^{\circ}$) observations to those derived
from occultation observations (Figure \ref{fig:temps} inset; Figure
\ref{fig:tc} bottom panel, blue dashes). The local
temperature maximum in 2012 at 350 km ($2.6\times10^{-2}$ mbar) is in
agreement with the altitudes of the detached
haze layer shortly after equinox as observed by Cassini and Voyager 2
\citep{rages_83, west_11}, following the disruption of Titan's
pole-to-pole circulation cell and the descent of the haze
layer. Contemporaneous Cassini/ISS images indicate that this feature
may be transient, or may be a secondary descending layer that
manifests during the
disappearance of the main detached haze layer below 300 km
\citep{seignovert_21}. Similarly, the inversion found in 2015
($L_S\sim68^{\circ}$) at 420
km ($6.8\times10^{-3}$ mbar) does not appear to correspond to a
stable, post-equinox haze layer through comparison with Cassini/ISS extinction
profiles, though the 2017 ($L_S\sim89^{\circ}$) feature at similar
altitudes aligns with the stable layer observed close to the end of
the Cassini mission \citep{west_18, seignovert_21}.

The absence of this feature in 2014 ($L_S\sim56^{\circ}$) and its subsequent re-emergence in
2015 and 2017 ($L_S\sim68$ and $89^{\circ}$) is in agreement with dynamical models of Titan's
atmosphere, marking the seasonal separation of haze due to Titan's
strengthening meridional circulation approaching solstice \citep{lebonnois_12, larson_15}. These altitudes are
similar to those found in the temperature inversions observed by
\citet{sicardy_99} in 1989, nearly one full Titan year prior. However,
as the morphology of the haze layer appears
inhomogeneous with latitude and longitude \citep{west_11, seignovert_21}, the nature of
the detached haze layer is difficult to quantify through the temperature
inversions discussed here due to the relatively large ALMA beam
footprint on Titan's disk. The ALMA resolution also smooths
  out small, latitudinally dependent vertical temperature
  perturbations, which complicates interpretations of these features
  and direct comparisons with previous observations. Further, while the measurement of 
temperature inversions was consistent across multiple 
retrievals (Figure \ref{fig:2015_temps}), the magnitude of these deviations are close to
the corresponding retrieval errors (Figure \ref{fig:temps}). As such, it must be cautioned that
the relation of these temperature features to the location of the
detached or transient haze
layer is tenuous, and somewhat dependent on the
confirmation of the latter from contemporaneous observations (e.g. those from
Cassini/ISS or occultation measurements).
Future ALMA observations at higher
spatial resolution may present the opportunity to confirm this
tentative connection and analyze the migration of the haze layer in a more meaningful way
than presented here. Those observations would enable more comprehensive studies of the
temporal and latitudinal evolution of Titan's haze after the end of the Cassini mission.

\subsection{Considerations and Future Work} \label{sec:cfw}
As the relative spatial and temporal scales enabled by the archival
ALMA observations do not allow for the full analysis of temperature
and HCN variability over relevant scales for Titan's upper atmosphere (e.g. a few Earth days,
across latitude and local time), a number of considerations must be
accounted for that constrain the interpretation of the results
presented here. First, we note that a direct comparison between our
ALMA results and those derived from HASI, INMS, and spacecraft occultation
measurements is difficult due to the limitations of vertical and
spatial resolution imposed by ground-based remote sensing
measurements. As a result, these observations are not sensitive to vertical structure variability (e.g. those seen by \citealp{fulchignoni_05},
\citealp{muller_06}, \citealp{snowden_13}) due to wave activity, which may
account for additional temperature variability at these altitudes on
the order 10 K \citep{strobel_06}. Similarly, the necessity to
simultaneously retrieve the vertical temperature and HCN profiles
from nadir and limb geometries somewhat obfuscates the direct assessment of
Titan's atmospheric state at individual locations, as longitudinal and diurnal effects in HCN
abundance and temperature may be introduced into the retrieved
profiles \citep{de_la_haye_08, cui_16}.

As addressed in previous
studies, complex atmospheric changes with latitude on Titan are not
well resolved due to the relatively large ALMA beam size \citep{thelen_18,
  thelen_19a}. This effect results in smoothing of vertical profiles
over large latitude
regions.
While previous analyses with Cassini
have shown no correlation between neutral thermospheric temperatures and latitude or
longitude \citep{snowden_13}, a full latitudinal map of mesospheric temperature
profiles would considerably facilitate comparisons with models of
Titan's atmosphere. Though beyond the scope of this work,
additional latitude points may be analyzed from 2016 and 2017
observations to characterize the variability of upper atmosphere
temperatures with latitude. Higher angular resolution observations will
also enable an improved assessment of the effects of wind shear and
dynamics on these temperature profiles. Higher cadence observations would allow for the
measurement of temperature variability in Titan's mesosphere to
quantify diurnal effects and those of Saturn's magnetosphere,
providing a comparison to measurements by Cassini/INMS in Titan's
thermosphere \citep{snowden_13, cui_16}. A study of the monthly
evolution of the mesospheric temperatures and winds during mid-2014 may be possible using
additional ALMA archive observations. Finally, future observations with ALMA may assist with the
development of updated spacecraft engineering models (e.g. for NASA's future
\textit{Dragonfly} mission), such as of \citet{yelle_97} for the
Huygens probe. 

\section{Conclusions} \label{sec:conc}
Through the analysis of yearly ALMA observations of Titan from 2012 to
2017 ($L_S\sim34$--$89^{\circ}$), we found large
variability in the temperature and HCN abundance of Titan's mesosphere
and thermosphere. The largest variation in mesospheric temperatures
occurs between 2014 and 2015 ($L_S\sim56$ and $68^{\circ}$), where the mesopause shifts dramatically
upwards by 250 km from $563 \pm 10$ km
to $812 \pm 42$ km, and cools from $153.7 \pm 4.0$ K to $134.9 \pm
3.4$ K (Figure \ref{fig:tc}). The observed
mesopause at both $\sim600$ and $\sim800$ km during this epoch
agrees with prior observations by Cassini and Keck II, and corroborates
the results of models of Titan's atmosphere; however, the large
variability observed here provides further challenges for modeling
Titan's thermal structure in the mesosphere and stratosphere. Smaller
temperature changes in the thermosphere and lower mesosphere may
be explained by previously observed perturbations in exogenic energy
sources, such as charged particle precipitation and interactions with
Saturn's magnetosphere, which occur on timescales shorter than the
observations analyzed here. Though latitudinal
  gradients in temperature were not derived from these observations,
  dynamical instabilities related to Titan's thermospheric jet
  (\citetalias{lellouch_19}; \citealp{cordiner_20}) may account for differences in the 2016
  and 2017 ($L_S\sim82$ and $89^{\circ}$)
  profiles. In the lower mesosphere, we retrieve small
inversions in the 2012, 2015, and 2017 temperature profiles that may
be indicative of the location of transient or detached haze layers,
located between $\sim350$--420 km following Titan's spring
equinox, similar to prior observations during this time period by
Cassini and one Titan year before by Voyager 2 and ground-based observations. 

Similarly large variability is found in Titan's HCN abundance, which
extends the observed thermospheric HCN by an order of magnitude, with
measurements from ALMA and Cassini now
comprising volume mixing ratios from
$\sim3\times10^{-5}$--$3\times10^{-3}$ above 900 km. While the
range in HCN abundance from year to year may account for
deviations in the
temperature profile from 2012 --- which is
relatively warm while the HCN abundance is fairly low --- and minor
temperature differences in the lower mesosphere and thermosphere,
the temperatures retrieved for 2014, 2015, and 2017 require energy
sources in addition to HCN cooling, solar UV/EUV heating, and wind shear. 
These sources may include variability in the abundances of Titan's
other atmospheric species produced through photochemistry, such as
C$_2$H$_2$ or C$_2$H$_6$, wave activity, or energy transfer from
Titan's lower atmosphere during the seasonal transition to northern
summer. 

Future observations with ALMA may help to further elucidate the
connection between Titan's upper atmospheric temperature and HCN
abundance through dedicated, high resolution observations over short
timescales (i.e. \textless10 Earth days) where diurnal differences and
those due to Titan's orbital position relative to the Saturnian plasma
environment are relevant. While the variability of Titan's
thermosphere was explored during the Cassini era,
mesospheric temperature differences with latitude, longitude, and
local time may be assessed through the unique capabilities of
ground-based (sub)millimeter facilities such as ALMA. In particular, these
observations are pertinent in the period around Titan's upcoming autumnal equinox in
2025, where similarly large variations in mesospheric temperature may
manifest. 

\section{Acknowledgements}
The authors wish to acknowledge the insightful discussion and comments
from E. Lellouch regarding ALMA data analysis and
radiative transfer modeling, and the helpful suggestions from two
anonymous reviewers that improved the quality of the final manuscript.

A.E.T. was funded by an appointment to the NASA Astrobiology
Postdoctoral Program at Goddard Space Flight Center, administered
by the USRA and ORAU through a contract
with NASA. CAN and MAC were supported for their contributions to this
work by NASA’s Solar System Observations (SSO) Program. NAT was funded
by the UK Science and Technology Facilities
Council. 

This paper makes use of the following ALMA data:
ADS/JAO.ALMA$\#$2012.0.00727.S, 2012.1.00635.S, 2013.1.00446.S,
2015.1.01023.S, and 2016.A.00014.S. ALMA is a partnership of ESO
(representing its member states), NSF (USA) and NINS (Japan), together
with NRC (Canada), MOST and ASIAA (Taiwan), and KASI (Republic of
Korea), in cooperation with the Republic of Chile. The Joint ALMA
Observatory is operated by ESO, AUI/NRAO and NAOJ. The National Radio
Astronomy Observatory is a facility of the National Science Foundation
operated under cooperative agreement by Associated Universities, Inc.

Some of the figures within this paper were produced using IDL colour-blind-friendly colour tables \citep[see][]{wright_17}.


\begin{thebibliography}{}
\expandafter\ifx\csname natexlab\endcsname\relax\def\natexlab#1{#1}\fi
\providecommand{\url}[1]{\href{#1}{#1}}
\providecommand{\dodoi}[1]{doi:~\href{http://doi.org/#1}{\nolinkurl{#1}}}
\providecommand{\doeprint}[1]{\href{http://ascl.net/#1}{\nolinkurl{http://ascl.net/#1}}}
\providecommand{\doarXiv}[1]{\href{https://arxiv.org/abs/#1}{\nolinkurl{https://arxiv.org/abs/#1}}}

\bibitem[{{Aboudan} {et~al.}(2008){Aboudan}, {Colombatti}, {Ferri}, \&
  {Angrilli}}]{aboudan_08}
{Aboudan}, A., {Colombatti}, G., {Ferri}, F., \& {Angrilli}, F. 2008, P$\&$SS,
  56, 573, \dodoi{10.1016/j.pss.2007.10.006}

\bibitem[{{Adriani} {et~al.}(2011){Adriani}, {Dinelli}, {L{\'o}pez-Puertas},
  {Garc{\'{\i}}a-Comas}, {Moriconi}, {D'Aversa}, {Funke}, \&
  {Coradini}}]{adriani_11}
{Adriani}, A., {Dinelli}, B.~M., {L{\'o}pez-Puertas}, M., {et~al.} 2011,
  Icarus, 214, 584, \dodoi{10.1016/j.icarus.2011.04.016}

\bibitem[{{Bell} {et~al.}(2014){Bell}, {Hunter Waite}, {Westlake}, {Bougher},
  {Ridley}, {Perryman}, \& {Mandt}}]{bell_14}
{Bell}, J.~M., {Hunter Waite}, J., {Westlake}, J.~H., {et~al.} 2014, Journal of
  Geophysical Research (Space Physics), 119, 4957, \dodoi{10.1002/2014JA019781}

\bibitem[{{Bell} {et~al.}(2011){Bell}, {Westlake}, \& {Waite}}]{bell_11}
{Bell}, J.~M., {Westlake}, J., \& {Waite}, J.~Hunter, J. 2011, GRL, 38, L06202,
  \dodoi{10.1029/2010GL046420}

\bibitem[{{Bell} {et~al.}(2010){Bell}, {Bougher}, {Waite}, {Ridley}, {Magee},
  {Mandt}, {Westlake}, {DeJong}, {Bar{\textendash}Nun}, {Jacovi}, {Toth}, \&
  {De La Haye}}]{bell_10}
{Bell}, J.~M., {Bougher}, S.~W., {Waite}, J.~H., {et~al.} 2010, JGR (Planets),
  115, E12002, \dodoi{10.1029/2010JE003636}

\bibitem[{{Bertucci} {et~al.}(2015){Bertucci}, {Hamilton}, {Kurth},
  {Hospodarsky}, {Mitchell}, {Sergis}, {Edberg}, \& {Dougherty}}]{bertucci_15}
{Bertucci}, C., {Hamilton}, D.~C., {Kurth}, W.~S., {et~al.} 2015, GRL, 42, 193,
  \dodoi{10.1002/2014GL062106}

\bibitem[{{Briggs}(1995)}]{briggs_95a}
{Briggs}, D.~S. 1995, PhD thesis, New Mexico Institute of Mining and Technology

\bibitem[{{Cordiner} {et~al.}(2020){Cordiner}, {Garcia-Berrios}, {Cosentino},
  {Teanby}, {Newman}, {Nixon}, {Thelen}, \& {Charnley}}]{cordiner_20}
{Cordiner}, M.~A., {Garcia-Berrios}, E., {Cosentino}, R.~G., {et~al.} 2020,
  ApJL, 904, L12, \dodoi{10.3847/2041-8213/abc688}

\bibitem[{{Cours} {et~al.}(2011){Cours}, {Burgalat}, {Rannou}, {Rodriguez},
  {Brahic}, \& {West}}]{cours_11}
{Cours}, T., {Burgalat}, J., {Rannou}, P., {et~al.} 2011, ApJL, 741, L32,
  \dodoi{10.1088/2041-8205/741/2/L32}

\bibitem[{Courtin {et~al.}(2011)Courtin, Swinyard, Moreno, Fulton, Lellouch,
  Rengel, \& Hartogh}]{courtin_11}
Courtin, R., Swinyard, B.~M., Moreno, R., {et~al.} 2011, A$\&$A, 536, L2

\bibitem[{{Coustenis} {et~al.}(2018){Coustenis}, {Jennings}, {Achterberg},
  {Bampasidis}, {Nixon}, {Lavvas}, {Cottini}, \& {Flasar}}]{coustenis_18}
{Coustenis}, A., {Jennings}, D.~E., {Achterberg}, R.~K., {et~al.} 2018, ApJL,
  854, L30, \dodoi{10.3847/2041-8213/aaadbd}

\bibitem[{{Coustenis} {et~al.}(2020){Coustenis}, {Jennings}, {Achterberg},
  {Lavvas}, {Bampasidis}, {Nixon}, \& {Flasar}}]{coustenis_20}
---. 2020, Icarus, 344, 113413, \dodoi{10.1016/j.icarus.2019.113413}

\bibitem[{{Creecy} {et~al.}(2019){Creecy}, {Li}, {Jiang}, {Nixon}, {West}, \&
  {Kenyon}}]{creecy_19}
{Creecy}, E.~C., {Li}, L., {Jiang}, X., {et~al.} 2019, GRL, 46, 13,649,
  \dodoi{10.1029/2019GL084833}

\bibitem[{{Creecy} {et~al.}(2021){Creecy}, {Li}, {Jiang}, {West}, {Fry},
  {Nixon}, {Kenyon}, \& {Seignovert}}]{creecy_21}
---. 2021, GRL, 48, e95356, \dodoi{10.1029/2021GL095356}

\bibitem[{{Cui} {et~al.}(2016){Cui}, {Cao}, {Lavvas}, \& {Koskinen}}]{cui_16}
{Cui}, J., {Cao}, Y.~T., {Lavvas}, P.~P., \& {Koskinen}, T.~T. 2016, ApJL, 826,
  L5, \dodoi{10.3847/2041-8205/826/1/L5}

\bibitem[{Cui {et~al.}(2009)Cui, Yelle, Vuitton, Waite~Jr., Kasprzak, Gell,
  Niemann, M{\"u}ller-Wodarg, Borggren, Fletcher, Patrick, Raaen, \&
  Magee}]{cui_09}
Cui, J., Yelle, R.~V., Vuitton, V., {et~al.} 2009, Icarus, 200, 581

\bibitem[{{de La Haye} {et~al.}(2008){de La Haye}, {Waite}, {Cravens},
  {Bougher}, {Robertson}, \& {Bell}}]{de_la_haye_08}
{de La Haye}, V., {Waite}, J.~H., {Cravens}, T.~E., {et~al.} 2008, Journal of
  Geophysical Research (Space Physics), 113, A11314,
  \dodoi{10.1029/2008JA013078}

\bibitem[{{Dobrijevic} {et~al.}(2021){Dobrijevic}, {Loison}, {Hue}, \&
  {Cavali{\'e}}}]{dobrijevic_21}
{Dobrijevic}, M., {Loison}, J.~C., {Hue}, V., \& {Cavali{\'e}}, T. 2021,
  Icarus, 364, 114477, \dodoi{10.1016/j.icarus.2021.114477}

\bibitem[{{Flasar} {et~al.}(2014){Flasar}, {Achterberg}, \&
  {Schinder}}]{flasar_14}
{Flasar}, F.~M., {Achterberg}, R.~K., \& {Schinder}, P.~J. 2014, Thermal
  structure of Titan's troposphere and middle atmosphere (Cambridge, UK:
  Cambridge University Press), 102

\bibitem[{Flasar {et~al.}(1981)Flasar, Samuelson, \& Conrath}]{flasar_81}
Flasar, F.~M., Samuelson, R.~E., \& Conrath, B.~J. 1981, Nature, 292, 293

\bibitem[{Flasar {et~al.}(2005)Flasar, Achterberg, Conrath, Gierasch, Kunde,
  Nixon, Bjoraker, Jennings, Romani, Simon-Miller, B{\'e}zard, Coustenis,
  Irwin, Teanby, Brasunas, Pearl, Segura, Carlson, Mamoutkine, Schinder,
  Barucci, Courtin, Fouchet, Gautier, Lellouch, Marten, Prag{\'e}, Vinatier,
  Strobel, Calcutt, Read, Taylor, Bowles, Samuelson, Orton, Spilker, Owen,
  Spencer, Showalter, Ferrari, Raulin, Edgington, Ade, \& Wishnow}]{flasar_05}
Flasar, F.~M., Achterberg, R.~K., Conrath, B.~J., {et~al.} 2005, Science, 308,
  975

\bibitem[{{Fomalont} {et~al.}(2014){Fomalont}, {van Kempen}, {Kneissl},
  {Marcelino}, {Barkats}, {Corder}, {Cortes}, {Hills}, {Lucas}, {Manning}, \&
  {Peck}}]{fomalont_14}
{Fomalont}, E., {van Kempen}, T., {Kneissl}, R., {et~al.} 2014, The Messenger,
  155, 19

\bibitem[{Fulchignoni {et~al.}(2005)Fulchignoni, Ferri, Angrilli, Ball,
  Bar-Nun, Barucci, Bettanini, Bianchini, Borucki, Colombatti, Coradini,
  Coustenis, Debei, Falkner, Fanti, Flamini, Gaborit, Grard, Hamelin, Harri,
  Hathi, Leese, Lehto, Lion~Stoppato, L{\'o}pez-Moreno, M{\"a}kinen, McDonnell,
  McKay, Molina-Cuberos, Neubauer, Pirronello, Rodrigo, Saggin, Schwingenschuh,
  Seiff, Sim{\~}oes, Svedhem, Tokano, Towner, Trautner, Withers, \&
  Zarnecki}]{fulchignoni_05}
Fulchignoni, M., Ferri, F., Angrilli, F., {et~al.} 2005, Nature, 438, 785,
  \dodoi{10.1038/nature04314}

\bibitem[{{Geballe} {et~al.}(2003){Geballe}, {Kim}, {Noll}, \&
  {Griffith}}]{geballe_03}
{Geballe}, T.~R., {Kim}, S.~J., {Noll}, K.~S., \& {Griffith}, C.~A. 2003, ApJL,
  583, L39, \dodoi{10.1086/368011}

\bibitem[{{Gordon} {et~al.}(2022){Gordon}, {Rothman}, {Hargreaves}, {Hashemi},
  {Karlovets}, {Skinner}, {Conway}, {Hill}, {Kochanov}, {Tan}, {Wcis{\l}o},
  {Finenko}, {Nelson}, {Bernath}, {Birk}, {Boudon}, {Campargue}, {Chance},
  {Coustenis}, {Drouin}, {Flaud}, {Gamache}, {Hodges}, {Jacquemart}, {Mlawer},
  {Nikitin}, {Perevalov}, {Rotger}, {Tennyson}, {Toon}, {Tran}, {Tyuterev},
  {Adkins}, {Baker}, {Barbe}, {Can{\`e}}, {Cs{\'a}sz{\'a}r}, {Dudaryonok},
  {Egorov}, {Fleisher}, {Fleurbaey}, {Foltynowicz}, {Furtenbacher}, {Harrison},
  {Hartmann}, {Horneman}, {Huang}, {Karman}, {Karns}, {Kassi}, {Kleiner},
  {Kofman}, {Kwabia-Tchana}, {Lavrentieva}, {Lee}, {Long}, {Lukashevskaya},
  {Lyulin}, {Makhnev}, {Matt}, {Massie}, {Melosso}, {Mikhailenko}, {Mondelain},
  {M{\"u}ller}, {Naumenko}, {Perrin}, {Polyansky}, {Raddaoui}, {Raston},
  {Reed}, {Rey}, {Richard}, {T{\'o}bi{\'a}s}, {Sadiek}, {Schwenke},
  {Starikova}, {Sung}, {Tamassia}, {Tashkun}, {Vander Auwera}, {Vasilenko},
  {Vigasin}, {Villanueva}, {Vispoel}, {Wagner}, {Yachmenev}, \&
  {Yurchenko}}]{gordon_22}
{Gordon}, I.~E., {Rothman}, L.~S., {Hargreaves}, R.~J., {et~al.} 2022, JQSRT,
  277, 107949, \dodoi{10.1016/j.jqsrt.2021.107949}

\bibitem[{Gurwell(2004)}]{gurwell_04}
Gurwell, M. 2004, ApJ, 616, L7

\bibitem[{{Hanel} {et~al.}(1981){Hanel}, {Conrath}, {Flasar}, {Kunde},
  {Maguire}, {Pearl}, {Pirraglia}, {Samuelson}, {Herath}, {Allison},
  {Cruikshank}, {Gautier}, {Gierasch}, {Horn}, {Koppany}, \&
  {Ponnamperuma}}]{hanel_81}
{Hanel}, R., {Conrath}, B., {Flasar}, F.~M., {et~al.} 1981, Science, 212, 192,
  \dodoi{10.1126/science.212.4491.192}

\bibitem[{{Hidayat} {et~al.}(1997){Hidayat}, {Marten}, {B{\'e}zard}, {Gautier},
  {Owen}, {Matthews}, \& {Paubert}}]{hidayat_97}
{Hidayat}, T., {Marten}, A., {B{\'e}zard}, B., {et~al.} 1997, Icarus, 126, 170,
  \dodoi{10.1006/icar.1996.5640}

\bibitem[{{H{\"o}gbom}(1974)}]{hogbom_74}
{H{\"o}gbom}, J.~A. 1974, A$\&$A Supplement, 15, 417

\bibitem[{{H{\"o}rst}(2017)}]{horst_17}
{H{\"o}rst}, S.~M. 2017, Journal of Geophysical Research (Planets), 122, 432,
  \dodoi{10.1002/2016JE005240}

\bibitem[{Irwin {et~al.}(2008)Irwin, Teanby, de~Kok, Fletcher, Howett, Tsang,
  Wilson, Calcutt, Nixon, \& Parrish}]{irwin_08}
Irwin, P. G.~J., Teanby, N.~A., de~Kok, R., {et~al.} 2008, J. Quant. Spec.
  Radiat. Transf., 109, 1136, \dodoi{10.1016/j.jqsrt.2007.11.006}

\bibitem[{{Jaeger}(2008)}]{jaeger_08}
{Jaeger}, S. 2008, in Astronomical Society of the Pacific Conference Series,
  Vol. 394, Astronomical Data Analysis Software and Systems XVII, ed. R.~W.
  {Argyle}, P.~S. {Bunclark}, \& J.~R. {Lewis}, 623

\bibitem[{{Kim} {et~al.}(2000){Kim}, {Geballe}, \& {Noll}}]{kim_00}
{Kim}, S.~J., {Geballe}, T.~R., \& {Noll}, K.~S. 2000, \icarus, 147, 588,
  \dodoi{10.1006/icar.2000.6481}

\bibitem[{{Kim} {et~al.}(2005){Kim}, {Geballe}, {Noll}, \& {Courtin}}]{kim_05}
{Kim}, S.~J., {Geballe}, T.~R., {Noll}, K.~S., \& {Courtin}, R. 2005, Icarus,
  173, 522, \dodoi{10.1016/j.icarus.2004.09.006}

\bibitem[{Koskinen {et~al.}(2011)Koskinen, Yelle, Snowden, Lavvas, Sandel,
  Capalbo, Benilan, \& West}]{koskinen_11}
Koskinen, T.~T., Yelle, R.~V., Snowden, D.~S., {et~al.} 2011, Icarus, 216, 507

\bibitem[{{Larson} {et~al.}(2015){Larson}, {Toon}, {West}, \&
  {Friedson}}]{larson_15}
{Larson}, E. J.~L., {Toon}, O.~B., {West}, R.~A., \& {Friedson}, A.~J. 2015,
  Icarus, 254, 122, \dodoi{10.1016/j.icarus.2015.03.010}

\bibitem[{{Lavvas} {et~al.}(2009){Lavvas}, {Yelle}, \& {Vuitton}}]{lavvas_09}
{Lavvas}, P., {Yelle}, R.~V., \& {Vuitton}, V. 2009, Icarus, 201, 626,
  \dodoi{10.1016/j.icarus.2009.01.004}

\bibitem[{Lebonnois {et~al.}(2012)Lebonnois, Burgalat, Rannou, \&
  Charnay}]{lebonnois_12}
Lebonnois, S., Burgalat, J., Rannou, P., \& Charnay, B. 2012, Icarus, 218, 707

\bibitem[{{Lellouch} {et~al.}(1990){Lellouch}, {Hunten}, {Kockarts}, \&
  {Coustenis}}]{lellouch_90}
{Lellouch}, E., {Hunten}, D.~M., {Kockarts}, G., \& {Coustenis}, A. 1990,
  Icarus, 83, 308, \dodoi{10.1016/0019-1035(90)90070-P}

\bibitem[{{Lellouch} {et~al.}(2019){Lellouch}, {Gurwell}, {Moreno}, {Vinatier},
  {Strobel}, {Moullet}, {Butler}, {Lara}, {Hidayat}, \&
  {Villard}}]{lellouch_19}
{Lellouch}, E., {Gurwell}, M.~A., {Moreno}, R., {et~al.} 2019, Nature
  Astronomy, 3, 614, \dodoi{10.1038/s41550-019-0749-4}

\bibitem[{{Li}(2015)}]{li_15}
{Li}, L. 2015, Scientific Reports, 5, 8239, \dodoi{10.1038/srep08239}

\bibitem[{{Liang} {et~al.}(2007){Liang}, {Yung}, \& {Shemansky}}]{liang_07}
{Liang}, M.-C., {Yung}, Y.~L., \& {Shemansky}, D.~E. 2007, ApJL, 661, L199,
  \dodoi{10.1086/518785}

\bibitem[{Lindal {et~al.}(1983)Lindal, Wood, Hotz, Sweetnam, Eshleman, \&
  Tyler}]{lindal_83}
Lindal, G.~F., Wood, G.~E., Hotz, H.~B., {et~al.} 1983, Icarus, 53, 348

\bibitem[{Loison {et~al.}(2015)Loison, H{\'e}brard, Dobrijevic, Hickson,
  Caralp, Hue, Gronoff, Venot, \& B{\'e}nilan}]{loison_15}
Loison, J.~C., H{\'e}brard, E., Dobrijevic, M., {et~al.} 2015, Icarus, 247,
  218, \dodoi{10.1016/j.icarus.2014.09.039}

\bibitem[{{Lorenz} {et~al.}(2014){Lorenz}, {Young}, \& {Ferri}}]{lorenz_14}
{Lorenz}, R.~D., {Young}, L.~A., \& {Ferri}, F. 2014, Icarus, 227, 49,
  \dodoi{10.1016/j.icarus.2013.08.025}

\bibitem[{{Magee} {et~al.}(2009){Magee}, {Waite}, {Mandt}, {Westlake}, {Bell},
  \& {Gell}}]{magee_09}
{Magee}, B.~A., {Waite}, J.~H., {Mandt}, K.~E., {et~al.} 2009, P$\&$SS, 57,
  1895, \dodoi{10.1016/j.pss.2009.06.016}

\bibitem[{Marten {et~al.}(2002)Marten, Hidayat, Biraud, \& Moreno}]{marten_02}
Marten, A., Hidayat, T., Biraud, Y., \& Moreno, R. 2002, Icarus, 158, 532

\bibitem[{{Math{\'e}} {et~al.}(2020){Math{\'e}}, {Vinatier}, {B{\'e}zard},
  {Lebonnois}, {Gorius}, {Jennings}, {Mamoutkine}, {Guandique}, \& {Vatant
  d'Ollone}}]{mathe_20}
{Math{\'e}}, C., {Vinatier}, S., {B{\'e}zard}, B., {et~al.} 2020, Icarus, 344,
  113547, \dodoi{10.1016/j.icarus.2019.113547}

\bibitem[{Molter {et~al.}(2016)Molter, Nixon, Cordiner, Serigano, Irwin,
  Teanby, Charnley, \& Lindberg}]{molter_16}
Molter, E.~M., Nixon, C.~A., Cordiner, M.~A., {et~al.} 2016, AJ, 152, 1,
  \dodoi{10.3847/0004-6256/152/2/42}

\bibitem[{{Moreno} {et~al.}(2005){Moreno}, {Marten}, \& {Hidayat}}]{moreno_05}
{Moreno}, R., {Marten}, A., \& {Hidayat}, T. 2005, A$\&$A, 437, 319,
  \dodoi{10.1051/0004-6361:20042117}

\bibitem[{M{\"u}ller {et~al.}(2001)M{\"u}ller, Thorwirth, Roth, \&
  Winnewisser}]{muller_01}
M{\"u}ller, H. S.~P., Thorwirth, S., Roth, D.~A., \& Winnewisser, G. 2001,
  A$\&$A, 370, L49, \dodoi{10.1051/0004-6361:20010367}

\bibitem[{{M{\"u}ller-Wodarg} {et~al.}(2006){M{\"u}ller-Wodarg}, {Yelle},
  {Borggren}, \& {Waite}}]{muller_06}
{M{\"u}ller-Wodarg}, I.~C.~F., {Yelle}, R.~V., {Borggren}, N., \& {Waite},
  J.~H. 2006, Journal of Geophysical Research (Space Physics), 111, A12315,
  \dodoi{10.1029/2006JA011961}

\bibitem[{{M{\"u}ller-Wodarg} {et~al.}(2008){M{\"u}ller-Wodarg}, {Yelle},
  {Cui}, \& {Waite}}]{muller_08}
{M{\"u}ller-Wodarg}, I.~C.~F., {Yelle}, R.~V., {Cui}, J., \& {Waite}, J.~H.
  2008, Journal of Geophysical Research (Planets), 113, E10005,
  \dodoi{10.1029/2007JE003033}

\bibitem[{{M{\"u}ller-Wodarg} {et~al.}(2000){M{\"u}ller-Wodarg}, {Yelle},
  {Mendillo}, {Young}, \& {Aylward}}]{muller_00}
{M{\"u}ller-Wodarg}, I.~C.~F., {Yelle}, R.~V., {Mendillo}, M., {Young}, L.~A.,
  \& {Aylward}, A.~D. 2000, JGR, 105, 20833, \dodoi{10.1029/2000JA000053}

\bibitem[{{Paubert} {et~al.}(1987){Paubert}, {Marten}, {Rosolen}, {Gautier}, \&
  {Courtin}}]{paubert_87}
{Paubert}, G., {Marten}, A., {Rosolen}, C., {Gautier}, D., \& {Courtin}, R.
  1987, in Bulletin of the American Astronomical Society, Vol.~19, 633

\bibitem[{{Porco} {et~al.}(2005){Porco}, {Baker}, {Barbara}, {Beurle},
  {Brahic}, {Burns}, {Charnoz}, {Cooper}, {Dawson}, {Del Genio}, {Denk},
  {Dones}, {Dyudina}, {Evans}, {Fussner}, {Giese}, {Grazier}, {Helfenstein},
  {Ingersoll}, {Jacobson}, {Johnson}, {McEwen}, {Murray}, {Neukum}, {Owen},
  {Perry}, {Roatsch}, {Spitale}, {Squyres}, {Thomas}, {Tiscareno}, {Turtle},
  {Vasavada}, {Veverka}, {Wagner}, \& {West}}]{porco_05}
{Porco}, C.~C., {Baker}, E., {Barbara}, J., {et~al.} 2005, Nature, 434, 159,
  \dodoi{10.1038/nature03436}

\bibitem[{{Rages} \& {Pollack}(1983)}]{rages_83}
{Rages}, K., \& {Pollack}, J.~B. 1983, Icarus, 55, 50,
  \dodoi{10.1016/0019-1035(83)90049-0}

\bibitem[{{Rannou} {et~al.}(2002){Rannou}, {Hourdin}, \& {McKay}}]{rannou_02}
{Rannou}, P., {Hourdin}, F., \& {McKay}, C.~P. 2002, Nature, 418, 853,
  \dodoi{10.1038/nature00961}

\bibitem[{{Rannou} {et~al.}(2004){Rannou}, {Hourdin}, {McKay}, \&
  {Luz}}]{rannou_04}
{Rannou}, P., {Hourdin}, F., {McKay}, C.~P., \& {Luz}, D. 2004, Icarus, 170,
  443, \dodoi{10.1016/j.icarus.2004.03.007}

\bibitem[{{Rengel} {et~al.}(2022){Rengel}, {Shulyak}, {Hartogh}, {Sagawa},
  {Moreno}, {Jarchow}, \& {Breitschwerdt}}]{rengel_22}
{Rengel}, M., {Shulyak}, D., {Hartogh}, P., {et~al.} 2022, A$\&$A, 658, A88,
  \dodoi{10.1051/0004-6361/202141422}

\bibitem[{{Rymer} {et~al.}(2009){Rymer}, {Smith}, {Wellbrock}, {Coates}, \&
  {Young}}]{rymer_09}
{Rymer}, A.~M., {Smith}, H.~T., {Wellbrock}, A., {Coates}, A.~J., \& {Young},
  D.~T. 2009, GRL, 36, L15109, \dodoi{10.1029/2009GL039427}

\bibitem[{{Schinder} {et~al.}(2020){Schinder}, {Flasar}, {Marouf}, {French},
  {Anabtawi}, {Barbinis}, {Fleischman}, \& {Achterberg}}]{schinder_20}
{Schinder}, P.~J., {Flasar}, F.~M., {Marouf}, E.~A., {et~al.} 2020, Icarus,
  345, 113720, \dodoi{10.1016/j.icarus.2020.113720}

\bibitem[{Schinder {et~al.}(2012)Schinder, Flasar, Marouf, French, McGhee,
  Kliore, Rappaport, Barbinis, Fleischman, \& Anabtawi}]{schinder_12}
Schinder, P.~J., Flasar, F.~M., Marouf, E.~A., {et~al.} 2012, Icarus, 221,
  1020, \dodoi{10.1016/j.icarus.2012.10.021}

\bibitem[{{Seignovert} {et~al.}(2017){Seignovert}, {Rannou}, {Lavvas}, {Cours},
  \& {West}}]{seignovert_17}
{Seignovert}, B., {Rannou}, P., {Lavvas}, P., {Cours}, T., \& {West}, R.~A.
  2017, Icarus, 292, 13, \dodoi{10.1016/j.icarus.2017.03.026}

\bibitem[{{Seignovert} {et~al.}(2021){Seignovert}, {Rannou}, {West}, \&
  {Vinatier}}]{seignovert_21}
{Seignovert}, B., {Rannou}, P., {West}, R.~A., \& {Vinatier}, S. 2021, ApJ,
  907, 36, \dodoi{10.3847/1538-4357/abcd3b}

\bibitem[{Serigano {et~al.}(2016)Serigano, Nixon, Cordiner, Irwin, Teanby,
  Charnley, \& Lindberg}]{serigano_16}
Serigano, J., Nixon, C.~A., Cordiner, M.~A., {et~al.} 2016, ApJ, 821, L8,
  \dodoi{10.3847/2041-8205/821/1/L8}

\bibitem[{{Sharkey} {et~al.}(2020){Sharkey}, {Teanby}, {Sylvestre}, {Mitchell},
  {Seviour}, {Nixon}, \& {Irwin}}]{sharkey_20}
{Sharkey}, J., {Teanby}, N.~A., {Sylvestre}, M., {et~al.} 2020, Icarus, 337,
  113441, \dodoi{10.1016/j.icarus.2019.113441}

\bibitem[{{Sharkey} {et~al.}(2021){Sharkey}, {Teanby}, {Sylvestre}, {Mitchell},
  {Seviour}, {Nixon}, \& {Irwin}}]{sharkey_21}
---. 2021, \icarus, 354, 114030, \dodoi{10.1016/j.icarus.2020.114030}

\bibitem[{{Shemansky} {et~al.}(2005){Shemansky}, {Stewart}, {West}, {Esposito},
  {Hallett}, \& {Liu}}]{shemansky_05}
{Shemansky}, D.~E., {Stewart}, A.~I.~F., {West}, R.~A., {et~al.} 2005, Science,
  308, 978, \dodoi{10.1126/science.1111790}

\bibitem[{{Sicardy} {et~al.}(1999){Sicardy}, {Ferri}, {Roques}, {Lecacheux},
  {Pau}, {Brosch}, {Nevo}, {Hubbard}, {Reitsema}, {Blanco}, {Carreira},
  {Beisker}, {Bittner}, {Bode}, {Bruns}, {Denzau}, {Nezel}, {Riedel},
  {Struckmann}, {Appleby}, {Forrest}, {Nicolson}, {Hollis}, \&
  {Miles}}]{sicardy_99}
{Sicardy}, B., {Ferri}, F., {Roques}, F., {et~al.} 1999, Icarus, 142, 357,
  \dodoi{10.1006/icar.1999.6219}

\bibitem[{{Sicardy} {et~al.}(2006){Sicardy}, {Colas}, {Widemann}, {Bellucci},
  {Beisker}, {Kretlow}, {Ferri}, {Lacour}, {Lecacheux}, {Lellouch}, {Pau},
  {Renner}, {Roques}, {Fienga}, {Etienne}, {Martinez}, {Glass}, {Baba},
  {Nagayama}, {Nagata}, {Itting-Enke}, {Bath}, {Bode}, {Bode}, {L{\"u}demann},
  {L{\"u}demann}, {Neubauer}, {Tegtmeier}, {Tegtmeier}, {Thom{\'e}}, {Hund},
  {deWitt}, {Fraser}, {Jansen}, {Jones}, {Schoenau}, {Turk}, {Meintjies},
  {Hernandez}, {Fiel}, {Frappa}, {Peyrot}, {Teng}, {Vignand}, {Hesler},
  {Payet}, {Howell}, {Kidger}, {Ortiz}, {Naranjo}, {Rosenzweig}, \&
  {Rapaport}}]{sicardy_06}
{Sicardy}, B., {Colas}, F., {Widemann}, T., {et~al.} 2006, JGR Planets, 111,
  E11S91, \dodoi{10.1029/2005JE002624}

\bibitem[{{Smith} {et~al.}(1982){Smith}, {Strobel}, {Broadfoot}, {Sandel},
  {Shemansky}, \& {Holberg}}]{smith_82}
{Smith}, G.~R., {Strobel}, D.~F., {Broadfoot}, A.~L., {et~al.} 1982, JGR, 87,
  1351, \dodoi{10.1029/JA087iA03p01351}

\bibitem[{{Snowden} \& {Higgins}(2021)}]{snowden_21}
{Snowden}, D., \& {Higgins}, A. 2021, Icarus, 354, 113929,
  \dodoi{10.1016/j.icarus.2020.113929}

\bibitem[{{Snowden} \& {Yelle}(2014)}]{snowden_14}
{Snowden}, D., \& {Yelle}, R.~V. 2014, Icarus, 228, 64,
  \dodoi{10.1016/j.icarus.2013.08.027}

\bibitem[{{Snowden} {et~al.}(2013){Snowden}, {Yelle}, {Cui}, {Wahlund},
  {Edberg}, \& {{\r{A}}gren}}]{snowden_13}
{Snowden}, D., {Yelle}, R.~V., {Cui}, J., {et~al.} 2013, Icarus, 226, 552,
  \dodoi{10.1016/j.icarus.2013.06.006}

\bibitem[{{Strobel}(2006)}]{strobel_06}
{Strobel}, D.~F. 2006, Icarus, 182, 251, \dodoi{10.1016/j.icarus.2005.12.015}

\bibitem[{{Sylvestre} {et~al.}(2020){Sylvestre}, {Teanby}, {Vatant d'Ollone},
  {Vinatier}, {B{\'e}zard}, {Lebonnois}, \& {Irwin}}]{sylvestre_20}
{Sylvestre}, M., {Teanby}, N.~A., {Vatant d'Ollone}, J., {et~al.} 2020, Icarus,
  344, 113188, \dodoi{10.1016/j.icarus.2019.02.003}

\bibitem[{{Teanby} {et~al.}(2019){Teanby}, {Sylvestre}, {Sharkey}, {Nixon},
  {Vinatier}, \& {Irwin}}]{teanby_19}
{Teanby}, N.~A., {Sylvestre}, M., {Sharkey}, J., {et~al.} 2019, GRL, 46, 3079,
  \dodoi{10.1029/2018GL081401}

\bibitem[{Teanby {et~al.}(2007)Teanby, Irwin, de~Kok, Vinatier, B{\'e}zard,
  Nixon, Flasar, Calcutt, Bowles, Fletcher, Howett, \& Taylor}]{teanby_07a}
Teanby, N.~A., Irwin, P. G.~J., de~Kok, R., {et~al.} 2007, Icarus, 186, 364

\bibitem[{Teanby {et~al.}(2008)Teanby, de~Kok, Irwin, Osprey, Vinatier,
  Gierasch, Read, Flasar, Conrath, Achterberg, B{\'e}zard, Nixon, \&
  Calcutt}]{teanby_08b}
Teanby, N.~A., de~Kok, R., Irwin, P. G.~J., {et~al.} 2008, J. Geophys. Res.,
  113, 1

\bibitem[{Teanby {et~al.}(2012)Teanby, Irwin, Nixon, de~Kok, Vinatier,
  Coustenis, Sefton-Nash, Calcutt, \& Flasar}]{teanby_12}
Teanby, N.~A., Irwin, P. G.~J., Nixon, C.~A., {et~al.} 2012, Nature, 491, 732

\bibitem[{{Teanby} {et~al.}(2017){Teanby}, {B{\'e}zard}, {Vinatier},
  {Sylvestre}, {Nixon}, {Irwin}, {de Kok}, {Calcutt}, \& {Flasar}}]{teanby_17}
{Teanby}, N.~A., {B{\'e}zard}, B., {Vinatier}, S., {et~al.} 2017, Nature
  Communications, 8, 1586, \dodoi{10.1038/s41467-017-01839-z}

\bibitem[{{Thelen} {et~al.}(2019{\natexlab{a}}){Thelen}, {Cordiner}, {Nixon},
  {Garcia Berrios}, {Charnley}, \& {Irwin}}]{thelen_agu_19}
{Thelen}, A.~E., {Cordiner}, M., {Nixon}, C.~A., {et~al.} 2019{\natexlab{a}},
  in {AGU Fall Meeting Abstracts}, Vol. 2019, P21C--06

\bibitem[{Thelen \& Molter(2018)}]{thelen_17_br}
Thelen, A.~E., \& Molter, E.~M. 2018, Mendeley Data, v1,
  \dodoi{10.17632/szbcb44s43.1}

\bibitem[{{Thelen} {et~al.}(2018){Thelen}, {Nixon}, {Chanover}, {Molter},
  {Cordiner}, {Achterberg}, {Serigano}, {Irwin}, {Teanby}, \&
  {Charnley}}]{thelen_18}
{Thelen}, A.~E., {Nixon}, C.~A., {Chanover}, N.~J., {et~al.} 2018, Icarus, 307,
  380, \dodoi{10.1016/j.icarus.2017.10.042}

\bibitem[{{Thelen} {et~al.}(2019{\natexlab{b}}){Thelen}, {Nixon}, {Chanover},
  {Cordiner}, {Molter}, {Teanby}, {Irwin}, {Serigano}, \&
  {Charnley}}]{thelen_19a}
---. 2019{\natexlab{b}}, Icarus, 319, 417, \dodoi{10.1016/j.icarus.2018.09.023}

\bibitem[{{Vervack} {et~al.}(2004){Vervack}, {Sandel}, \&
  {Strobel}}]{vervack_04}
{Vervack}, R.~J., {Sandel}, B.~R., \& {Strobel}, D.~F. 2004, Icarus, 170, 91,
  \dodoi{10.1016/j.icarus.2004.03.005}

\bibitem[{Vinatier {et~al.}(2015)Vinatier, B{\'e}zard, Lebonnois, Teanby,
  Achterberg, Gorius, Mamoutkine, Guandique, Jolly, Jennings, \&
  Flasar}]{vinatier_15}
Vinatier, S., B{\'e}zard, B., Lebonnois, S., {et~al.} 2015, Icarus, 250, 95

\bibitem[{{Vinatier} {et~al.}(2020){Vinatier}, {Math{\'e}}, {B{\'e}zard},
  {Vatant d'Ollone}, {Lebonnois}, {Dauphin}, {Flasar}, {Achterberg},
  {Seignovert}, {Sylvestre}, {Teanby}, {Gorius}, {Mamoutkine}, {Guandique}, \&
  {Jennings}}]{vinatier_20}
{Vinatier}, S., {Math{\'e}}, C., {B{\'e}zard}, B., {et~al.} 2020, A$\&$A, 641,
  A116, \dodoi{10.1051/0004-6361/202038411}

\bibitem[{{Vuitton} {et~al.}(2019){Vuitton}, {Yelle}, {Klippenstein},
  {H{\"o}rst}, \& {Lavvas}}]{vuitton_19}
{Vuitton}, V., {Yelle}, R.~V., {Klippenstein}, S.~J., {H{\"o}rst}, S.~M., \&
  {Lavvas}, P. 2019, Icarus, 324, 120, \dodoi{10.1016/j.icarus.2018.06.013}

\bibitem[{Vuitton {et~al.}(2007)Vuitton, Yelle, \& McEwan}]{vuitton_07}
Vuitton, V., Yelle, R.~V., \& McEwan, M.~J. 2007, Icarus, 191, 722,
  \dodoi{10.1016/j.icarus.2007.06.023}

\bibitem[{{West} {et~al.}(2011){West}, {Balloch}, {Dumont}, {Lavvas}, {Lorenz},
  {Rannou}, {Ray}, \& {Turtle}}]{west_11}
{West}, R.~A., {Balloch}, J., {Dumont}, P., {et~al.} 2011, JGRL, 38, L06204,
  \dodoi{10.1029/2011GL046843}

\bibitem[{{West} {et~al.}(2018){West}, {Seignovert}, {Rannou}, {Dumont},
  {Turtle}, {Perry}, {Roy}, \& {Ovanessian}}]{west_18}
{West}, R.~A., {Seignovert}, B., {Rannou}, P., {et~al.} 2018, Nature Astronomy,
  2, 495, \dodoi{10.1038/s41550-018-0434-z}

\bibitem[{{Westlake} {et~al.}(2011){Westlake}, {Bell}, {Waite}, {Johnson},
  {Luhmann}, {Mandt}, {Magee}, \& {Rymer}}]{westlake_11}
{Westlake}, J.~H., {Bell}, J.~M., {Waite}, J.~H., J., {et~al.} 2011, Journal of
  Geophysical Research (Space Physics), 116, A03318,
  \dodoi{10.1029/2010JA016251}

\bibitem[{{Willacy} {et~al.}(2016){Willacy}, {Allen}, \& {Yung}}]{willacy_16}
{Willacy}, K., {Allen}, M., \& {Yung}, Y. 2016, ApJ, 829, 79,
  \dodoi{10.3847/0004-637X/829/2/79}

\bibitem[{{Wright}(2017)}]{wright_17}
{Wright}, P. 2017, {ColourBlind: A Collection of Colour-blind-friendly Colour
  Tables},  Zenodo, \dodoi{10.5281/zenodo.840393}

\bibitem[{{Yelle}(1991)}]{yelle_91}
{Yelle}, R.~V. 1991, ApJ, 383, 380, \dodoi{10.1086/170796}

\bibitem[{{Yelle} {et~al.}(2006){Yelle}, {Borggren}, {de la Haye}, {Kasprzak},
  {Niemann}, {M{\"u}ller-Wodarg}, \& {Waite}}]{yelle_06}
{Yelle}, R.~V., {Borggren}, N., {de la Haye}, V., {et~al.} 2006, Icarus, 182,
  567, \dodoi{10.1016/j.icarus.2005.10.029}

\bibitem[{{Yelle} {et~al.}(2008){Yelle}, {Cui}, \&
  {M{\"u}ller-Wodarg}}]{yelle_08}
{Yelle}, R.~V., {Cui}, J., \& {M{\"u}ller-Wodarg}, I.~C.~F. 2008, Journal of
  Geophysical Research (Planets), 113, E10003, \dodoi{10.1029/2007JE003031}

\bibitem[{{Yelle} \& {Griffith}(2003)}]{yelle_03}
{Yelle}, R.~V., \& {Griffith}, C.~A. 2003, Icarus, 166, 107,
  \dodoi{10.1016/S0019-1035(03)00218-5}

\bibitem[{{Yelle} {et~al.}(2014){Yelle}, {Snowden}, \&
  {M{\"u}ller-Wodarg}}]{yelle_14}
{Yelle}, R.~V., {Snowden}, D.~S., \& {M{\"u}ller-Wodarg}, I.~C.~F. 2014,
  {Titan's upper atmosphere: thermal structure, dynamics, and energetics}
  (Cambridge, UK: Cambridge University Press), 322

\bibitem[{{Yelle} {et~al.}(1997){Yelle}, {Strobell}, {Lellouch}, \&
  {Gautier}}]{yelle_97}
{Yelle}, R.~V., {Strobell}, D.~F., {Lellouch}, E., \& {Gautier}, D. 1997, in
  ESA Special Publication, Vol. 1177, Huygens: Science, Payload and Mission,
  ed. A.~{Wilson}, 243

\end{thebibliography}
\end{document}